\documentclass[12pt,a4paper]{article}

\usepackage[a4paper]{geometry}

\expandafter\let\csname equation*\endcsname\relax
\expandafter\let\csname endequation*\endcsname\relax
\usepackage{amsmath, amsfonts}

\numberwithin{equation}{section}

\usepackage{hyperref}
\usepackage{color}
\usepackage{graphicx}
\usepackage{mathtools}
\usepackage{caption}
\usepackage{subcaption}

\renewcommand{\refeq}[1]{(\ref{#1})}
\newcommand{\reffig}[1]{Fig.\;\ref{#1}}
\newcommand{\refsec}[1]{\S\ref{#1}}
\newcommand{\integers}{\mathbb{Z}}
\newcommand{\complex}{\mathbb{C}}
\newcommand{\real}{\mathbb{R}}

\newcommand{\jth}[1]{\vartheta_{#1}}
\renewcommand{\inf}{\infty}

\renewcommand{\sc}[1]{\text{sc}\left(#1\right)}

\newcommand{\deriv}[2]{\frac{\mathrm{d}#1}{\mathrm{d}#2}}

\newcommand{\shift}{\Delta}

\newcommand{\Det}[1]{\text{Det}\left[#1\right]}
\newcommand{\p}{\vspace{6pt}\noindent}
\newcommand{\jump}{\vspace{2pt}}

\usepackage{amsxtra}
    \usepackage{amstext}
    \usepackage{amssymb}
    \usepackage{latexsym}
    \usepackage{graphicx}
\usepackage{graphics}

\newcommand{\abs}[1]{\left|#1\right|}

\advance\textheight by \topskip
\makeatletter

\@addtoreset{equation}{section}
\def\section{\@startsection {section}{1}{\z@}{-8.5ex plus -1ex minus
 -.2ex}{3.3ex plus .2ex}{\large\bf}}
\def\subsection{\@startsection{subsection}{2}{\z@}{-3.25ex plus
 -1ex minus -.2ex}{1.5ex plus .2ex}{\bf}}
\def\subsubsection{\@startsection{subsubsection}{3}{\z@}{-3.25ex plus%
 -1ex minus -.2ex}{1.5ex plus .2ex}{\sl}}

\begin{document}

\begin{titlepage}
\vspace*{-2cm}
\begin{flushright}
\end{flushright}

\vspace{0.3cm}

\begin{center}
{\Large {\bf }} \vspace{1cm} {\Large {\bf Type I integrable defects and finite-gap solutions for\\
 \vspace{5pt}KdV and sine-Gordon models}}\\
\vspace{1cm} {\large  E.\ Corrigan\footnote{\noindent E-mail: {\tt
ec9@york.ac.uk}} and
R.\ Parini\footnote{\noindent E-mail: {\tt rp910@york.ac.uk}} \\
\vspace{0.5cm}

{\em Department of Mathematics\\ University of York\\  \vspace{.25cm} York YO10 5DD, U.K.}} \\

\vspace{2cm} {\bf{ABSTRACT}}
\end{center}
\p The main purpose of this paper is to extend results, which have been obtained previously to describe the classical scattering of solitons with integrable defects of type I, to include the much larger and intricate collection of finite-gap solutions defined in terms of generalised theta functions. In this context, it is generally not feasible to adopt a direct approach, via ans\"atze for the fields to either side of the defect tuned to satisfy the defect sewing conditions. Rather, essential use is made of the fact that the defect sewing conditions themselves are intimately related to B\"acklund transformations in order to set up a strategy to enable the calculation of the field on one side by suitably transforming the field on the other side. The method is implemented using Darboux transformations and illustrated in detail for the sine-Gordon and KdV models.  An exception, treatable by both methods, indirect and direct, is provided by the genus 1 solutions. These can be expressed in terms of Jacobi elliptic functions, which satisfy a number of useful identities of relevance to this problem. There are new features to the solutions obtained in the finite-gap context but, in all cases, if a (multi)soliton limit is taken within the finite-gap solutions previously known results are recovered.

\vfill
\end{titlepage}


\section{Introduction}

It has been discovered that many systems of nonlinear integrable partial differential field equations allow for the introduction of certain special types of discontinuity, or \lq defect', that do not appear to violate integrability. These are described by \lq sewing' conditions that relate the field (or, more typically, its time or space derivatives) on one side of the defect to similar quantities on the other side. For example, this has been shown to be possible within a number of systems, including sine-Gordon and other affine Toda field theories \cite{Bowcock2004}, the nonlinear Schr{\"o}dinger, Korteweg-de Vries (KdV), mKdV \cite{Corrigan2006} and complex sine-Gordon \cite{Bowcock2008}. For a systematic approach, see also \cite{Caudrelier2008}. A characteristic property of these defects is their ability to store energy and momentum and thus exchange both quantities with the fields. This allows the total energy and momentum of the system to be conserved (alongside other, higher spin, charges). Indeed, there is evidence that these special defects, with sewing conditions that can be described within a Lagrangian formulation of the field equations, provide a classical description of purely transmitting defects of a type that had been considered much earlier within integrable quantum field theory \cite{Delf94, Konik97, Bowcock2005}.

\p Some special solutions to integrable systems with a \lq type I' defect (a defect that does not carry any additional degrees of freedom of its own) have been described using solitons, which are well-known special localised solutions that also carry a well-defined energy and momentum.
For both  sine-Gordon \cite{Bowcock2005} and KdV \cite{Corrigan2006} it was found that a single soliton passing through a defect is adjusted by a phase shift, in a sense that will be described further below, but the velocity of the soliton is unchanged.  In sine-Gordon, as a consequence of scattering with the defect the topological charge of a soliton might change sign, or it might be captured by the defect, or, for KdV the soliton might emerge as a singularity. Exactly what happens very much depends on the choice of defect parameter and the velocity of the approaching soliton.
Beyond these phase-shifted solutions, there is also the possibility that a defect carrying topological charge together with energy and momentum might create a soliton. However, this process, if it could occur would need additional initial data to specify the time of creation \cite{Bowcock2005, Corrigan2006}. On the other hand, while this may appear strange from a classical perspective, the interpretation of this process within quantum field theory as the decay of a resonance seems quite natural.
Originally, all these features were found by direct substitutions of an ansatz to solve the equations of motion of the fields to either side of the defect and match the defect sewing conditions.

\p More generally, sine-Gordon, KdV and many other integrable systems are known to possess quasi-periodic (finite-gap) solutions that depend upon choices of branch points on a hyperelliptic Riemann surface. These solutions are formulated in terms of Riemann theta functions (see, for example, \cite{Kotlarov1976,Dubrovin1981,Mumford,AGANIE}). Because of their quasi-periodic nature it would not be expected that energy-momentum or other conserved quantities should have well-defined total values but it is quite natural nevertheless to ask what effect an inserted defect might have and what the consequences might be of sewing together solutions of this more general type using the defect conditions. Indeed, there may be circumstances in which an average energy or momentum could be defined and be useful. It is also worth recalling that multi-soliton solutions can be obtained as limits of the finite-gap solutions in which pairs of branch points coalesce \cite{Mumford}. This means that the already known results concerning solitons should emerge as limits taken within the moduli spaces of  more general solutions. It is the central objective of this paper to find novel finite-gap solutions to sine-Gordon and KdV in the presence of type I defects.

\p Solutions constructed using the data from genus 1 Riemann surfaces can be expressed in terms of Jacobi elliptic functions. For these it is feasible to adopt a direct approach by making  explicit ansatze for the fields on either side of the defect then arranging them to satisfy the sewing conditions directly. This will be accomplished explicitly  in section 4 below. However, for more general solutions, using data from higher genus surfaces, this approach is not practicable. Instead, it will be useful to make use of an observation made some time ago in \cite{Bowcock2004}. It was noted that the type I defect sewing conditions for sine-Gordon, KdV and other types of integrable equations resemble a B{\"a}cklund transformation applied at a particular point in space rather than over the full line. A more formal description of the relationship between defects in space, or time, and B{\"a}cklund transformations has been developed in \cite{Doikou2016} but here the observation will be used as a tool to generate solutions to the sine-Gordon and KdV field equations in the presence of a type I defect.

\p The essential idea is straightforward and can be summarised as follows. Given a solution $u(x,t)$, where the field to the left of the defect is $u(x<x_D,t)$, one may first perform a B{\"a}cklund transformation for all $x,t$ to find $v(x,t)$.
Then the defect equations at the point $x=x_D$ are satisfied by $u(x_D,t)$, $v(x_D,t)$ and their derivatives, and the field to the right of the defect will be $v(x>x_D,t)$.
Thus, in essence, solutions are sought for which the defect is a manifestation of a B{\"a}cklund transformation, connecting a field to its B{\"a}cklund transformed self across the defect.

\p The B{\"a}cklund transformation is implemented on the level of the Lax pair using a Darboux transformation \cite{Matveev1991,Gu2005} and it is demonstrated that \lq one-to-two' soliton solutions, where the initial state of the defect has a discontinuity, can be derived using this method. Additionally, it is found that the phase-shifted solutions may be recovered as limits of the one-to-two soliton solutions where the time at which the additional soliton is created is taken to $\pm\inf$.

\p The method is then extended to generate solutions that satisfy the defect sewing equations on a finite-gap background of arbitrary genus.  It will be shown that if the field to the left of the defect is a finite-gap solution then the field to the right may contain a soliton, created by the defect at an undetermined time, on a similar finite-gap background of the same genus.
In the appropriate limits the phase-shifted one soliton or one-to-two soliton solutions to the defect equations can be obtained from this finite-gap solution.
In the sine-Gordon case  a purely phase-shifted genus 1 solution, found by direct substitution of an ansatz into the defect equations, is matched with a particular case of the finite-gap solution found by using a Darboux transformation.

\p The paper is arranged as follows. Section 2 summarises the main known features of classical defect-soliton scattering, which are subsequently used to illustrate the approach that will be adopted, while section 3 reviews the main features of the finite-gap solutions to sine-Gordon, in preparation for the main novel results contained in sections 4 and 5. Sections 6 and 7 use corresponding techniques to find similar finite-gap solutions to KdV that can be matched across a defect. Section 8 contains concluding remarks and suggestions for future directions to explore. Some of the detailed calculations are elaborate and those relating to soliton limits, or partial soliton limits in which a genus $g+1$ solution becomes a genus $g$ solution plus a soliton, are relegated to appendices A and B, respectively.
\newpage

\section{Soliton solutions to the defect equations for sine-Gordon}

The sine-Gordon equation (with all dimensionful parameters removed by scaling) is
\begin{equation}\label{SG}
	\quad w_{tt} - w_{xx} + \sin\,w = 0.
\end{equation}
With a  type I defect at the point $x=x_D$, it is convenient to denote the fields in the regions to the left and right of the defect by $w(x<x_D,t)=u(x,t)$ and $w(x>x_D,t)=v(x,t)$, so that each satisfies  \refeq{SG} in its own domain and at the point $x=x_D$ the sewing conditions take the form \cite{Bowcock2004},
\begin{subequations} \label{SGdefecteqs}
\begin{align}
	\quad &u_x = v_t - \sigma \sin\left(\frac{u+v}{2}\right) - \sigma^{-1} \sin\left(\frac{u-v}{2}\right) \label{SGdefecteq1}\\
	\quad &v_x = u_t + \sigma \sin\left(\frac{u+v}{2}\right) - \sigma^{-1} \sin\left(\frac{u-v}{2}\right). \label{SGdefecteq2}
\end{align}
\end{subequations}
The parameter $\sigma$ appearing in \refeq{SGdefecteqs} is called the defect parameter and has an important role to play.

\subsection{Purely phase-shifted soliton solutions}
\label{sec:SGPhaseShifted}

It is convenient to express a  single soliton solution $u(x,t)$ to the sine-Gordon equation by,
\begin{equation} \label{SGsoliton}
	e^{iu/2} = \frac{1+iE}{1-iE},\quad
	E = e^{ax+bt+c},\quad
	a=\cosh\theta,\quad
	b=-\sinh\theta,\quad
\end{equation}
where $\theta$ is the rapidity of the soliton and $\theta>0$ corresponds to a soliton moving in the positive direction along the $x$ axis.  A corresponding antisoliton solution may be obtained by sending $c \rightarrow c+i\pi$ so that $E\rightarrow -E$.

\p If a soliton of the form \refeq{SGsoliton} is moving towards the defect \refeq{SGdefecteqs} from the left then, as in \cite{Bowcock2004}, a solution to \refeq{SGdefecteqs} can be found by assuming, since the matching must be achievable for all time, the field on the right to be a similar solution but with a phase shift.
Solving for the phase shift using \refeq{SGdefecteqs} gives,
\begin{equation} \label{SGv1to1_cont}
	e^{iv/2} = \frac{1+i\delta E}{1-i\delta E},\qquad
	\delta = \coth\left(\frac{\eta-\theta}{2}\right),
\end{equation}
where the defect parameter $\sigma=e^{-\eta}>0$ and $E$ is as in \refeq{SGsoliton}.
Although $\sigma > 0$ is assumed the corresponding solution with $\sigma \rightarrow -\sigma$ can be found by also exchanging $u$ and $v$ since this simply swaps \refeq{SGdefecteq1} and \refeq{SGdefecteq2}.

\p A purely phase-shifted solution distinct from the one given in \cite{Bowcock2004} can be found by noting that although the sine-Gordon equation is invariant under $v \rightarrow v + 2 \pi n$ for $n \in \integers$, the defect equations \refeq{SGdefecteqs} are only invariant under $v \rightarrow v + 4 \pi n$.
So it is possible to write down a distinct solution with $u$ again just \refeq{SGsoliton} but with the ansatz for $v$ shifted by $2\pi$.
Then using \refeq{SGdefecteqs} to solve for the phase shift shows that it is the inverse of the previous case,
\begin{equation}\label{SGv1to1_discont}
	e^{iv/2} = -\frac{1+i\delta^{-1} E}{1-i\delta^{-1} E}
\end{equation}
In the first case, \refeq{SGv1to1_cont}, the initial condition for the fields does not permit the defect to store any energy-momentum above the ground state configuration, $u=v=0$, and the soliton is phase-shifted backwards and delayed. In the second case, \refeq{SGv1to1_discont}, the initial configuration for the fields contains a $2\pi$ discontinuity at the defect, which therefore stores the energy and momentum equal to that of a soliton of rapidity $\eta$ and the soliton is phase-shifted forwards.
In both cases if $\eta > \theta$ the soliton remains a soliton while if $\eta < \theta$ $\delta$ is negative and the soliton emerges as an antisoliton.
If $\theta=\eta$ then in either case the solution requires $\exp(iv/2)=-1$ and the final field configuration is a $2\pi$ singularity at the defect. However, the interpretations in the two cases are different:  for \refeq{SGv1to1_cont} the soliton is infinitely phase-shifted backwards so that nothing ever emerges from the defect while for \refeq{SGv1to1_discont} the soliton is infinitely phase-shifted forwards and hence should be considered to have already emerged a long time in the past and lies at $x=+\infty$. Given that the initial discontinuity in the generic case ($\theta\ne\eta$) could be thought of as a \lq hidden' soliton the latter is consistent with the known fact that there is a repulsion between solitons and there is no classical solution with two solitons of equal rapidity. In all cases, energy and momentum are conserved while being exchanged over time with the defect.

\subsection{Soliton Creation}
\label{sec:SGSolCreation}

Another family of solutions to the defect equations, discussed in \cite{Bowcock2005}, involves soliton creation.
If $u$ is the one soliton solution \refeq{SGsoliton} then a solution for the other side of the defect, $v$, is given by
\begin{equation} \label{SGv1to2}
\begin{gathered}
	e^{iv/2} = \frac{1+i\delta E_\theta \pm iE_\eta \mp \delta^{-1} E_\theta E_\eta}{1-i\delta E_\theta \mp iE_\eta \mp \delta^{-1} E_\theta E_\eta} \\
	E_\theta  = \exp\left[\cosh\theta\, x - \sinh\theta\,t - x_{\theta}\right] \\
	E_\eta  = \exp\left[\cosh\eta \, x - \sinh\eta\,t - x_{\eta}\right]
\end{gathered}
\end{equation}
The existence of this solution can be derived as a consequence of the ability of the defect to destroy a soliton together with a symmetry of the defect equations.
To see this take $u$ to be a two soliton solution where one of the solitons has rapidity equal to $\eta$.
Since solitons are individually affected by the defect it is clear from \refsec{sec:SGPhaseShifted} that the soliton with rapidity $\eta$ will be annihilated leaving the field on the other side of defect as \refeq{SGv1to1_discont}.
But if $(u,v)$ solves \refeq{SGdefecteqs} then so too does $(\tilde u, \tilde v)=(v \pm 2\pi,u)$.
Therefore, a two-to-one soliton solution to the defect equations is transformed under this symmetry to a one-to-two soliton solution where a soliton of rapidity $\eta$ is created by the defect.
The initial position for the created soliton, $x_\eta$, and the choice of $\pm$ (which corresponds to the created soliton being a kink or antikink), is not fixed by the given $u$ or defect parameter $\sigma=e^{-\eta}$.

\p It has been previously noted \cite{Bowcock2005} that the phase shift $\delta$ experienced by a single soliton passing through the defect is the square root of the total phase shift experienced by the same soliton being overtaken by a soliton of rapidity $\eta$.
 However, the phase-shifted solutions \refeq{SGv1to1_cont} or \refeq{SGv1to1_discont} can be alternatively and directly obtained from \refeq{SGv1to2} by taking the limits $x_\eta \rightarrow \infty$ or $x_\eta \rightarrow -\infty$, respectively.

\p In the context of quantum field theory the free choice of $x_\eta$ is reflected by the fact that the transmission matrix associated with a soliton passing through the defect has a pole at a certain complex rapidity that can be interpreted as an unstable soliton-defect bound sate with a finite decay width \cite{Bowcock2005}.
Classically, one might imagine that in a physical situation there would be knowledge of $u$ and $v$ at some initial time that would allow the solution to be fixed.
In \cite{Bowcock2005} it was shown that this is the case if the field at the defect is initially continuous (mod $4\pi$) as $t\rightarrow -\infty$.  With $u$ being the one soliton solution \refeq{SGsoliton} it is in this case energetically impossible for an additional soliton to be created and the only solution for $v$ is \refeq{SGv1to1_cont}.
If instead the field has a discontinuity of $2\pi$ (mod $4\pi$) at the defect then a soliton could be created but the presence of the discontinuity by itself is not sufficient to determine the time at which the new soliton would be released.

\subsection{Defects and B{\"a}cklund transformations}
\label{sec:DefectsAndBT}

It is a remarkable fact that when the defect equations \refeq{SGdefecteqs} are applied over all $x$ instead of a single point they have the form of a B{\"a}cklund transformation \cite{Bowcock2004}.
In the context of integrable PDEs B{\"a}cklund transformations are used in conjunction with Bianchi's permutability theorem to generate multisoliton solutions \cite{Drazin1989}.

\p In fact, the $u$ \refeq{SGsoliton} and $v$ \refeq{SGv1to2} that constitute the one-to-two soliton solution of the defect equations are related to each other by a B{\"a}cklund transformation.
That is to say that \refeq{SGsoliton} and \refeq{SGv1to2} actually solve \refeq{SGdefecteqs} for \textit{all} $x$ as well as at the point $x=x_D$ where the defect is located.
This might have been anticipated since \refeq{SGsoliton} and \refeq{SGv1to2} are completely independent of the defect's position so they would have to solve \refeq{SGdefecteqs} for any choice of $x_D$.

\p The role of the defect then appears to be to connect a given field in $x<x_D$ to its B{\"a}cklund transformed field in $x>x_D$ with B{\"a}cklund parameter equal to the defect parameter.
This suggests a systematic method of constructing solutions to the defect equations by taking a solution to sine-Gordon on the full line, $u(x,t)$ and performing a B{\"a}cklund transformation  to find $v(x,t)$.  A solution satisfying the defect  sewing equations at the point $x=x_D$ is then simply $u$ for $x<x_D$ and $v$ for $x>x_D$. This is the method adopted in \refsec{sec:SGFiniteGapDefects} to derive finite-gap solutions to the defect equations for arbitrary genus.

\section{Finite-gap solutions to sine-Gordon}
\label{sec:SGFiniteGap}

In this section some known facts concerning finite-gap solutions will be reviewed both to introduce notation and for completeness.

\p The finite-gap solutions are written in terms of Riemann theta functions
\begin{equation}
	\theta(z,B) = \sum_{n\in\integers^g} e^{\frac{1}{2}n\cdot Bn + n\cdot z}, \quad
	z\in\complex^g
\end{equation}
where $B$ is a symmetric $g\times g$ matrix with negative real part known as the Riemann matrix and $g\ge 1$ is an integer denoting the genus of a Riemann surface.
Riemann theta functions are quasi-periodic,  satisfying
\begin{equation}
	\theta(z+2\pi ip+Bq,B) = e^{-\frac{1}{2}q\cdot Bq -q\cdot z} \theta(z,B) \qquad p,q \in \integers^g.
\end{equation}
In terms of these special functions, the finite-gap solutions to the sine-Gordon equation on the full line are \cite{Kotlarov1976, AGANIE},
\begin{equation} \label{SGfinitegap}
	e^{i u(x,t)/2} = \frac{\theta(iVx+iWt+D, B)}{\theta(iVx+iWt+D+i\pi, B)}, \quad
	D = x_0 + \frac{i\pi}{2} \kappa + i\pi \varepsilon,
\end{equation}
where  to ensure that the solution is real
\begin{equation}
	x_0,\ \kappa,\ \epsilon \in \real^g,
	\quad
	\varepsilon_i = 0 \text{ or } 1,
	\quad
	\kappa_i =
	\begin{cases}
	    1, & \text{if } p_i, q_i \in \real \\
	    0, & \text{if } p_i = \bar q_i
	\end{cases}\ ,\quad i=1\dots g,
\end{equation}
and $W$, $V$ and $B$ are defined by a choice of branch points $p_i, q_i$ on a hyperelliptic Riemann surface.
For sine-Gordon this is the surface consisting of the points $(\mu,\lambda)$ such that,
\begin{equation} \label{SG_RS}
	\mu^2 = \lambda \prod\limits_{i=1}^{g}(\lambda-p_i)(\lambda-q_i)
\end{equation}
where each pair of branch points, $p_i, q_i$ may either be real with $p_i < q_i < 0$ or complex conjugates $p_i = \bar q_i$.
It is possible to have conjugate pairs of branch points whose midpoints are on the positive real axis (for example, the $g=1$ case is detailed in \cite{Forest82}) but for simplicity only conjugate pairs for which $p_i+q_i<0$ will be considered explicitly although the results of \refsec{sec:SGFiniteGapDefects} are expected also  to apply more generally.

\begin{figure}
	\center
	\includegraphics[width=15cm]{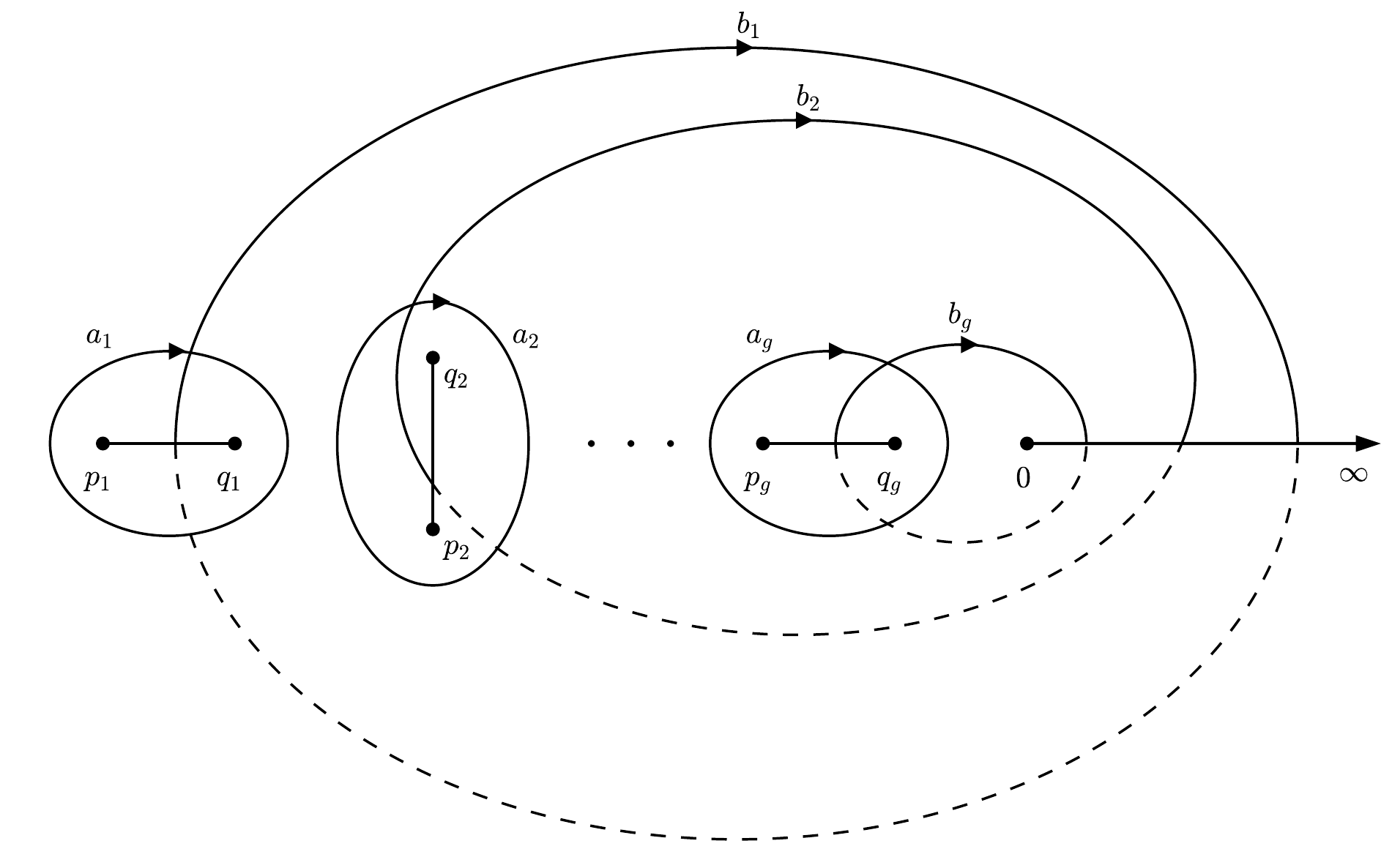}
	\caption{
	The basis of cycles $a_i$, $b_i$ where the parts of the cycles on the upper sheet have solid lines and parts on the lower sheet have dashed lines.
	The solid lines between the branch points $(p_i,q_i)$ and $(0,\infty)$ denote branch cuts.
	}
	\label{SGHomology}
\end{figure}

\p The surface \refeq{SG_RS} is two-sheeted and the branch cuts are chosen to lie between each pair of points $p_i, q_i$ and on the interval $(0,\infty)$, as shown in \reffig{SGHomology}.

\p The upper sheet is defined by the condition that on the upper sheet $\mu(\lambda)>0$ for $\lambda$ on the upper side of the cut from $0$ to $\infty$.
It will be computationally useful to explicitly implement the upper sheet as specified in \cite{Frauendiener2004}.
Let $\sqrt{z}$ denote the principal square root with a branch cut along $(-\infty, 0)$, and Arg$(z)$  be the associated argument function with ${-\pi < \text{Arg}(z) \leq \pi}$.
Then for the branch cut between branch points $p$ and $q$,
\begin{equation} \label{upperSheet}
\sqrt[(p,q)]{z} =
\begin{cases}
    \phantom{-} \sqrt{z}, & \text{if Arg$(q-p) \leq $ Arg$(z) < $ Arg$(q-p) + 2\pi$},\\
    -\sqrt{z}, & \text{otherwise}.
\end{cases}
\end{equation}
The upper sheet is therefore implemented for sine-Gordon by
\begin{equation}\label{upperSheeta}
	\sqrt[(0,\infty)]{\lambda}
	\prod\limits_{i=1}^{g}
	\sqrt[(p_i,q_i)]{\lambda-p_i}
	\sqrt[(p_i,q_i)]{\lambda-q_i}.
\end{equation}

\p A basis of cycles $a_i,\,b_i$ on the Riemann surface is chosen as shown in \reffig{SGHomology}.
The $a_i$ cycle encircles the $i$th branch cut clockwise on the upper sheet.
The $b_i$ cycle begins on the upper sheet, moves clockwise through the $(0,\infty)$ branch cut, changing to the lower sheet, and returns to the upper sheet by passing through the $i$th branch cut.

\p The holomorphic differentials on this surface are \cite{Baker}
\begin{equation} \label{holomorphicDiff}
	\omega_i = C_{ij} \frac{\lambda^{j-1}}{\mu} d\lambda, \quad i,j = {1,\cdots,g},
\end{equation}
where the normalization constants $C_{ij}$ are defined by the condition
\begin{equation} \label{holomorphicNormalization}
	\oint_{a_i} \omega_j = 2\pi i \delta_{ij}.
\end{equation}
Then the Riemann matrix is
\begin{equation} \label{RiemannMatrix}
	B_{ij} = \oint_{b_i} \omega_j.
\end{equation}

\p Sometimes it will be convenient to write the periods of the $a_i$ and $b_i$ cycles as line integrals between branch points (as in, for example, \cite{Frauendiener2014,Bobenko2011})
\begin{equation} \label{PeriodsAsLineIntegrals}
	\oint_{a_i} \omega_j = 2\int_{p_i}^{q_i} \omega_j \qquad
	\oint_{b_i} \omega_j = 2\sum_{k=i}^{g-1} \int_{q_k}^{p_{k+1}} \omega_j + 2\int_{q_g}^{0}\omega_j,
\end{equation}
where the line integrals are taken to be over the upper sheet.  Another useful relation comes from the fact that the sum of all $a_i$ cycles is homologous to the positively oriented contour around the cut $(0,\infty)$ \cite{Bobenko2011}. Then, taking into account the normalisation \refeq{holomorphicNormalization}, the following holds:
\begin{equation} \label{aCycleSum}
	\int_{\infty}^0 \omega_j
	= \sum_{i=1}^g \int_{p_i}^{q_i} \omega_j
	= \sum_{i=1}^g i\pi \delta_{ij}
	= i\pi,
\end{equation}
where the integrals are over the upper sheet.

\p To define $V$ and $W$ the Abelian integrals of the second kind are needed \cite{AGANIE}
\begin{equation}
	\Omega_1(P) = \int_{0}^P d\Omega_1(P), \qquad
	\Omega_2(P) = \int_{\infty}^P d\Omega_2(P),
\end{equation}
with singularities of the form,
\begin{subequations}
 \label{SGsecondKind}
 \begin{align}
	\Omega_1 & \rightarrow k_1 + O(k_1^{-1}), \qquad k_1 \rightarrow \infty, \qquad k_1=\sqrt{\lambda} \\
	\Omega_2 & \rightarrow k_2 + O(k_2^{-1}), \qquad k_2 \rightarrow \infty, \qquad k_2=1/\sqrt{\lambda},
 \end{align}
\end{subequations}
for the local parameters $k_i$.
Explicitly, the corresponding differentials are
\begin{subequations}
 \label{SGsecondKindDiff}
 \begin{align}
	d\Omega_1 &= \frac{\lambda^g}{2\mu} d\lambda + \sum_{i=1}^g \alpha_i \omega_i, \\
	d\Omega_2 &= -\frac{\sqrt{\Lambda}}{2\lambda\mu} d\lambda + \sum_{i=1}^g \beta_i \omega_i, \qquad \Lambda = \prod_{i=1}^g p_i q_i, \label{SGsecondKindDiff2}
 \end{align}
\end{subequations}
where the constants $\alpha_i, \beta_i$ are fixed by the normalization condition
\begin{equation} \label{SGsecondNormalization}
	\oint_{a_i} d\Omega_1 = 0, \qquad
	\oint_{a_i} d\Omega_2 = 0
\end{equation}
and the $g$-dimensional periods are
\begin{equation}
	(B_{1})_i = \oint_{b_i} d\Omega_1, \qquad
	(B_{2})_i = \oint_{b_i} d\Omega_2.
\end{equation}
Then the coefficients of $x,t$ appearing in \refeq{SGfinitegap} are given by
\begin{equation}
	V = \frac{B_1 - B_2}{4}, \qquad
	W = \frac{B_1 + B_2}{4}.
\end{equation}

\p The values of $B_1$ and $B_2$ can be  written conveniently in terms of the holomorphic differential normalization constants $C_{ij}$ using a Riemann bilinear relation.
Specifically, for any integral of the first kind (holomorphic) $\Omega$ and any integral of the second kind $\tilde\Omega$ with a pole at $P_0$, each with $a_k$ and $b_k$ periods $A_k,B_k$ and $\tilde A_k, \tilde B_k$, respectively, there is the relation \cite[eq(2.4.13)]{AGANIE}
\begin{equation} \label{RiemannBilinear}
	\sum_k \left(A_k \tilde B_k - \tilde A_k B_k\right) =
	\frac{2\pi i}{(n-1)!}\left. \frac{d^{n-1}}{dz^{n-1}} \Omega(z)\right|_{z=z_0},
\end{equation}
where the local parameter $z$ in the neighbourhood of $P_0$ is chosen so that
\begin{equation}
	d\tilde\Omega = [(z-z_0)^{-n} + O(1)]dz, \quad n>1.
\end{equation}
For the differentials \refeq{SGsecondKind},
\begin{align}
	d\Omega_1 &= [z^{-2} + O(1)]dz, \quad z=-\lambda^{-1/2} \\
	d\Omega_2 &= [z^{-2} + O(1)]dz, \quad z=-{\lambda}^{1/2},
\end{align}
which, using the normalization conditions \refeq{holomorphicNormalization} and \refeq{SGsecondNormalization}, gives
\begin{equation} \label{SGSecondB}
	(B_1)_i =  2 C_{ig}, \qquad
	(B_2)_i = -\frac{2C_{i1}}{\sqrt{\prod_{k=1}^g p_k q_k}}.
\end{equation}

\section{Phase-shifted genus 1 solutions for sine-Gordon with a defect}

\subsection{Genus 1 solutions on the full line}

Before considering the case of arbitrary genus, it is possible in the genus 1 case to find phase-shifted solutions by direct substitution of an ansatz into the defect equations in the same way that the phase-shifted soliton solution was found in \cite{Bowcock2004}. To achieve this it will be convenient within this section to use the notation of Jacobi theta functions:
\begin{align*}
	& \jth{1}(z,B)= -\jth{2}(z+i\pi,B) &
	& \jth{2}(z,B)= \sum_{n=-\inf}^{\inf} \exp\left(\frac{B}{2}\left(n+\frac{1}{2}\right)^2 + z\left(n+\frac{1}{2}\right)\right) & \\
	& \jth{3}(z,B)= \theta(z,B) &
	& \jth{4}(z,B)= \theta(z+i\pi,B) &
\end{align*}
The abbreviations $\jth{k}(z)=\jth{k}(z,B)$ and $\jth{k}=\jth{k}(0,B)$ will also be useful.
Note that the more common definition of Jacobi theta functions, $\jth{k}(u,q)$, used in \cite{Whittaker1962} and implemented in Mathematica, is related to the notation used here by $\jth{k}(u=z/(2i),q=\exp(B/2))$.

\p For sine-Gordon in the genus 1 case the unnormalized a-period for the holomorphic differential can be expressed simply in terms of branch points \cite{Enolski2008} \cite[\S 13.20(7)]{Bateman1953}
\begin{equation} \label{SGEllipticA}
	A
	:= \oint_{a} \frac{d\lambda}{\mu}
	=- \frac{2\pi \jth{3}^2}{\sqrt{-p_1}}
	= -\frac{2\pi \jth{4}^2}{\sqrt{-q_1}}
	= -\frac{2\pi \jth{2}^2}{\sqrt{q_1 - p_1}}.
\end{equation}

\p The sign of $A$ depends on which sheet the $a$ cycle is taken to be on so it is worthwhile to check that \refeq{SGEllipticA} corresponds to our choice of $a$ by deriving the first equality of \refeq{SGEllipticA}.
Using \refeq{PeriodsAsLineIntegrals} the a-period can be written as $A = 2\int_{p_1}^{q_1} d\lambda/\mu$ where the path of the integral is on the upper sheet.
This integral can be put in the form of standard elliptic integrals with the substitution $\lambda=t^2(q_1-p_1)+p_1$.  Ensuring always that square roots remain on the upper sheet defined by \refeq{upperSheeta}, it follows that, for $p_1<q_1<0$ or $q_1=\bar p_1$ with $\text{Re}[q_1]<0$,
\begin{equation} \label{AEllipticK}
	A = -\frac{4}{\sqrt{-p_1}} K \left(1-\frac{q_1}{p_1}\right),
\end{equation}
where $K(m)$ is the complete elliptic integral of the first kind
\begin{equation}
	K(m) = \int_0^1 \frac{dt}{\sqrt{1-t^2}\sqrt{1-m t^2}}.
\end{equation}
One of the relations between periods of elliptic functions and theta functions is $2K(1-q_1/p_1)=\pi \jth{3}^2$ \cite[\S 22.302]{Whittaker1962} and hence \refeq{AEllipticK} verifies the first equality of \refeq{SGEllipticA}.

\p Using \refeq{SGEllipticA} the normalization constant may be written as,
\begin{equation} \label{SGEllipticC}
	C = 2i\pi A^{-1} = \frac{(p_1 q_1)^{1/4}}{i \jth{3} \jth{4}},
\end{equation}
so that \refeq{SGSecondB} becomes
\begin{equation} \label{VW}
	B_1 = 2C = - \frac{2i (p_1 q_1)^{1/4}}{\jth{3} \jth{4}},
	\qquad
	B_2 = \frac{2}{C \jth{3}^2 \jth{4}^2} = \frac{2i (p_1 q_1)^{-1/4}}{\jth{3} \jth{4}}.
\end{equation}

\subsection{Genus 1 phase-shifted solution}
\label{sec:SG1GensuPhaseShifted}

The purely phase-shifted genus 1 solution to the defect equations can be obtained by inserting the ansatz
\begin{equation}\label{defectjacobi}
	e^{iu/2} = \frac{\jth{3}(z)}{\jth{4}(z)},
	\quad
	e^{iv/2} = \frac{\jth{3}(z+\Delta)}{\jth{4}(z+\Delta)},
	\quad
	z = iVx+iWt+D,
\end{equation}
directly into \refeq{SGdefecteqs} and finding the phase shift that correctly sews the two fields together at the defect.
To do this the derivatives are computed  using \cite[\S21.6]{Whittaker1962}
\begin{equation}\label{derivformula}
	2i \deriv{}{z} \left[\frac{\jth{3}(z)}{\jth{4}(z)}\right] = -\jth{2}^2 \, \frac{\jth{1}(z)\jth{2}(z)}{\jth{4}(z)^2},
\end{equation}
 to find a pair of equations both linear in ${\jth{1}(z+\Delta)\jth{2}(z+\Delta)/\jth{4}^2(z+\Delta)}$ and quadratic in ${\jth{3}(z+\Delta)/\jth{4}(z+\Delta)}$.

\p Solving for ${\jth{3}(z+\Delta)/\jth{4}(z+\Delta)}$ and using \refeq{VW} together with the well-known relations between the squares of Jacobi theta functions \cite[\S21.2]{Whittaker1962}
\begin{equation} \label{ThetaRelations}
\begin{aligned}
	\jth{1}^2(z) \jth{2}^2 &= \jth{4}^2(z)\jth{3}^2 - \jth{3}^2(z)\jth{4}^2 \quad
	&& \jth{3}^2(z) \jth{2}^2 = \jth{2}^2(z)\jth{3}^2 + \jth{1}^2(z)\jth{4}^2 \\
	\jth{2}^2(z) \jth{2}^2 &= \jth{3}^2(z)\jth{3}^2 - \jth{4}^2(z)\jth{4}^2 \quad
	&& \jth{4}^2(z) \jth{2}^2 = \jth{1}^2(z)\jth{3}^2 + \jth{2}^2(z)\jth{4}^2
\end{aligned}
\end{equation}
it is found that
\begin{equation}\label{defectth3th4}
	\frac{\jth{3}(z+\Delta)}{\jth{4}(z+\Delta)}
	=
	\frac{\jth{1}(z) \jth{2}(z) \jth{2}^2 C \sigma
	\pm
	\jth{3}(z) \jth{4}(z)
	\sqrt{(C^2 \jth{3}^4 + \sigma^2)(C^2 \jth{4}^4 + \sigma^2)}}
	{C^2 \jth{3}^2 \jth{4}^2 \jth{4}^2(z) + \jth{3}^2(z) \sigma^2},
\end{equation}
which, using \refeq{defectjacobi}, gives the field to the right of the defect.  In order to isolate the phase change $\Delta$ it is useful to compare \refeq{defectth3th4} with the addition formula for theta functions:
\begin{equation}\label{additionth3th4}
	\frac{\jth{3}(z+\Delta)}{\jth{4}(z+\Delta)}
	=
	\frac{\jth{4}}{\jth{3}}\;
	\frac{\jth{1}(z)\jth{2}(z)\jth{1}(\Delta)\jth{2}(\Delta) - \jth{3}(z)\jth{4}(z)\jth{3}(\Delta)\jth{4}(\Delta)}{\jth{1}^2(z)\jth{1}^2(\Delta) - \jth{4}^2(z)\jth{4}^2(\Delta)}.
\end{equation}
 The next step is to find a relationship between $\Delta$ and the parameters $C$ and $\sigma$ to ensure the equality of \refeq{defectth3th4} and \refeq{additionth3th4} for all $z$. Noting first, using \refeq{ThetaRelations}, that any theta function can be written in terms of any other pair, it is helpful to eliminate $\jth{1}(z)$ and $\jth{2}(z)$ before equating coefficients of $\jth{3}(z)$ and $\jth{4}(z)$.
This can be done by equating \refeq{defectth3th4} and \refeq{additionth3th4}, rearranging to find
\begin{multline} \label{EliminateTheta12}
	\jth{1}(z)\jth{2}(z)\left(\jth{2}^2 C \sigma F(z) - \jth{1}(\Delta)\jth{2}(\Delta) G(z) \right)
	= \\
	- \jth{3}(z)\jth{4}(z)\left(\jth{3}(\Delta)\jth{4}(\Delta)G(z) \pm F(z) \sqrt{(C^2 \jth{3}^4 + \sigma^2)(C^2 \jth{4}^4 + \sigma^2)}\right),
\end{multline}
with
\begin{align*}
	F(z) &= \frac{\jth{3}}{\jth{4}} \left(\jth{1}^2(\Delta)\jth{1}^2(z) - \jth{4}^2(\Delta)\jth{4}^2(z)\right) \\
	G(z) &= C^2 \jth{3}^2\jth{4}^2\jth{4}^2(z) + \sigma^2 \jth{3}^2(z),
\end{align*}
and then squaring.
After making use of \refeq{ThetaRelations} to eliminate $\jth{1}^2(z)$, $\jth{2}^2(z)$, $\jth{3}(\Delta)$ and $\jth{4}(\Delta)$, \refeq{EliminateTheta12} becomes a polynomial in $\jth{3}(z)$ and $\jth{4}(z)$ with coefficients depending on $\jth{1}(\Delta)$, $\jth{2}(\Delta)$, $C$, $\sigma$ and the theta constants $\jth{i}$.
With the use of computer algebra it can be shown that for all of these coefficients to vanish it is required that
\begin{equation} \label{jacobidefectphase}
	\frac{\jth{1}(\Delta)}{\jth{2}(\Delta)} = -\frac{\sigma}{\jth{3}\jth{4}C}.
\end{equation}
Note that since the zeros of Jacobi theta functions are are given by \cite[\S 21.12]{Whittaker1962}
\begin{align}
	\jth{1}(z) = 0 &\iff z = 2\pi in + Bm, \\
	\jth{2}(z) = 0 &\iff z = 2\pi in + Bm + i\pi,\\
	\jth{3}(z) = 0 &\iff z = 2\pi in + Bm + i\pi + B/2, \\
	\jth{4}(z) = 0 &\iff z = 2\pi in + Bm + B/2,
\end{align}
where $n,m \in \integers$, it is clear that the theta constants $\jth{2}$, $\jth{3}$, $\jth{4}$ are non-zero and that $\jth{1}(\Delta)$ and $\jth{2}(\Delta)$ cannot both be zero.
This observation eliminates other possible constraints on the parameters that might cause the coefficients of $\jth{3}(z)$, $\jth{4}(z)$ to vanish.

\p Assuming \refeq{jacobidefectphase} the expression obtained for ${\jth{1}(z+\Delta)\jth{2}(z+\Delta)/\jth{4}^2(z+\Delta)}$ from direct substitution of the ansatz into the defect equations can be equated with its corresponding addition formula.

\begin{figure}
\begin{subfigure}[t]{0.485\textwidth}
  \centering
  \hspace*{-15pt}
  \includegraphics[width=1.07\linewidth]{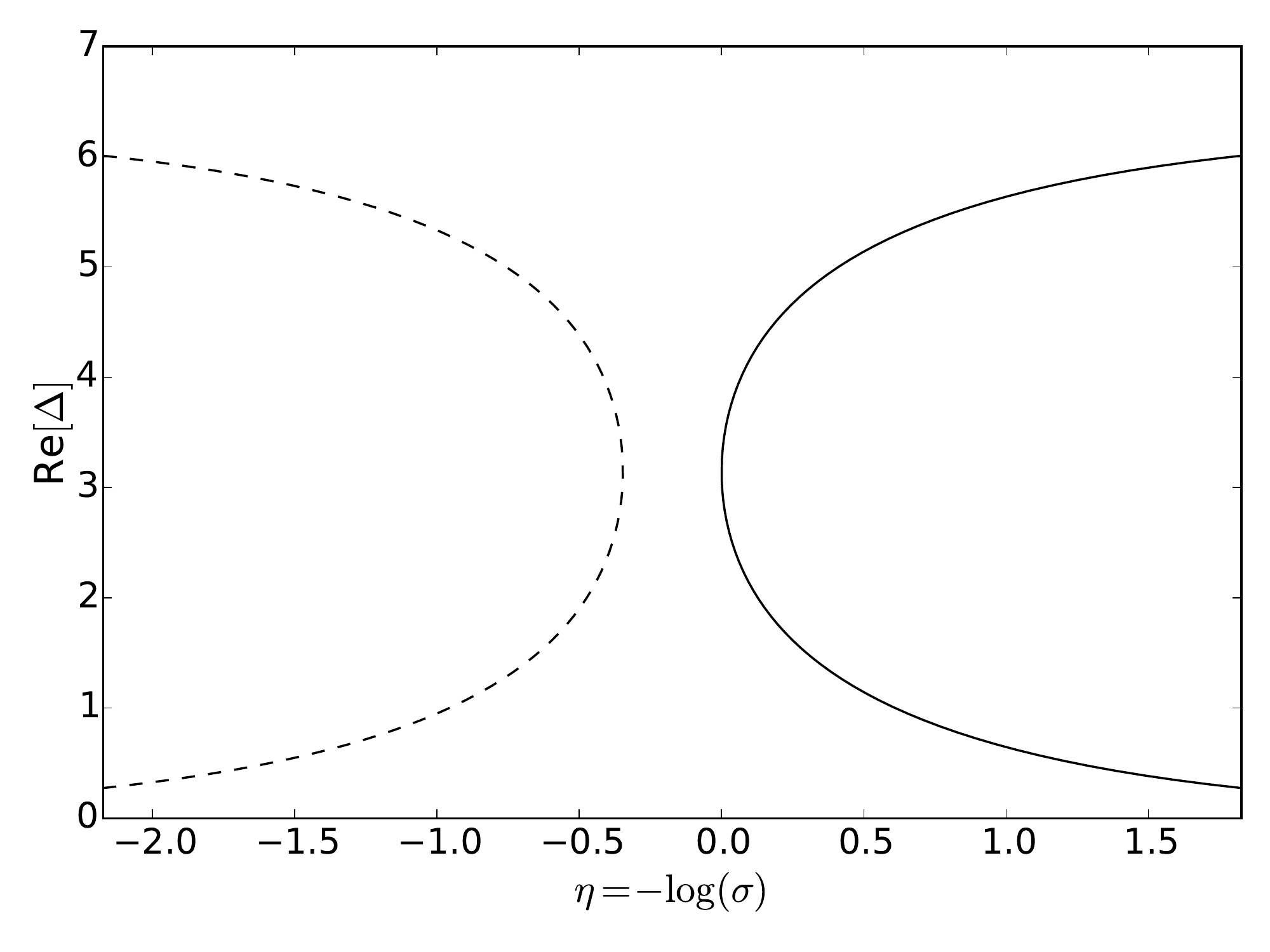}
  \caption{$p_1=-2$, $q_1=-1$}
\end{subfigure}
\hfill
\begin{subfigure}[t]{0.485\textwidth}
  \centering
  \hspace*{-15pt}
  \includegraphics[width=1.07\linewidth]{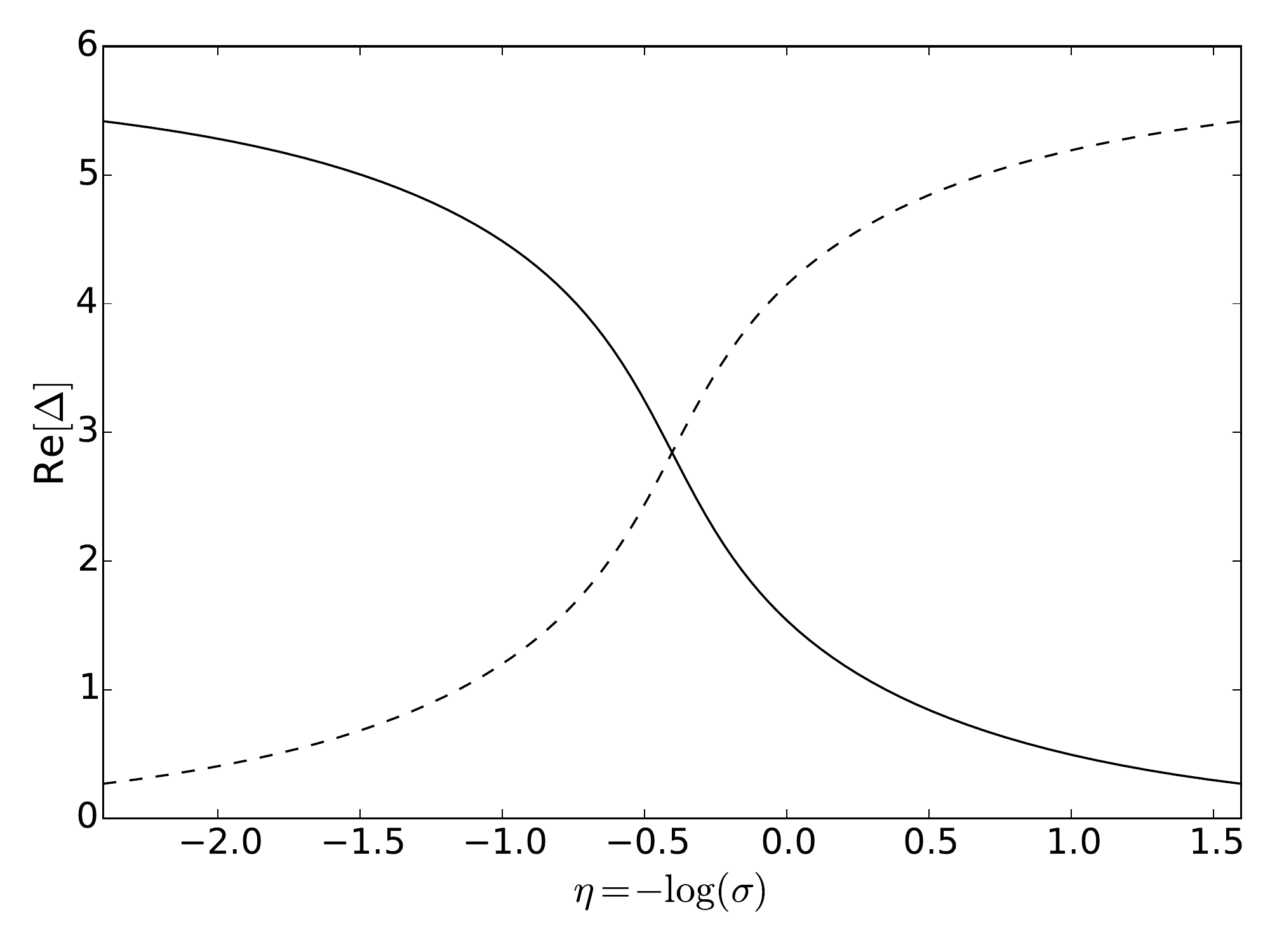}
  \caption{$p_1=-2-i$, $q_1=-2+i$}
\end{subfigure}
\caption{
For (a) real and (b) conjugate branch points values of the phase shift $\Delta$ are plotted that correspond to the two distinct solutions for the field $v$ given in \refeq{defectjacobi}.
Where the line is solid $\Delta$ is real while the dashed line indicates that Im$[\Delta]=\pi$.  If Im$[\Delta]=\pi$ then the sign of the phase-shifted field $v$ will be reversed compared to $u$.
}
\label{DeltaFig}
\end{figure}

\p Using \refeq{jacobidefectphase} it is now apparent that the $\pm$ in \refeq{defectth3th4} is a consequence of the relationship between $\jth{3}(\Delta)\jth{4}(\Delta)$ and $\jth{1}(\Delta)$ and $\jth{2}(\Delta)$ only being determined up to a sign by the square relations \refeq{ThetaRelations}.
This indicates that there are two distinct (mod $4\pi$) purely phase-shifted solutions in the genus 1 case, just as there were in the soliton case.
In fact, if $\Delta$ solves \refeq{jacobidefectphase} then so does $-\Delta+B$,
\begin{equation}
	\frac{\jth{1}(-\Delta+B)}{\jth{2}(-\Delta+B)} = -\frac{\jth{1}(-\Delta)}{\jth{2}(-\Delta)} = \frac{\jth{1}(\Delta)}{\jth{2}(\Delta)}
\end{equation}
since $\jth{1}(z)$ is an odd function of $z$ and $\jth{2}(z)$, $\jth{3}(z)$, $\jth{4}(z)$ are even.
This gives two distinct solutions for the field $v$ for a given value of $\Delta$,
\begin{equation} \label{SGEllipticV}
	e^{iv/2} =  \frac{\jth{3}(z+\Delta)}{\jth{4}(z+\Delta)}, \qquad
	e^{iv/2} = -\frac{\jth{3}(z-\Delta)}{\jth{4}(z-\Delta)}
\end{equation}
It is also true that $\Delta + 2\pi in + 2Bm$ will solve \refeq{jacobidefectphase} if $\Delta$ does for any $n,m\in\integers$ but this does not lead to a different $v$.
In \reffig{DeltaFig}  it is noted in some examples how the phase shift in the genus 1 case varies as a function of the defect parameter for real and imaginary cuts.

\p One feature of \refeq{jacobidefectphase}, which agrees with the soliton case, is that the effect of the defect disappears in the limit where $\sigma\rightarrow 0$.  In this limit \refeq{jacobidefectphase} implies that $\jth{1}(\Delta)=0$ in which case $v=u$ or $v=u+2\pi$.  In the soliton case, if $\sigma\rightarrow 0$ then \refeq{SGv1to1_cont} becomes $v=u$ and \refeq{SGv1to1_discont} gives $v=u+2\pi$.

\p As a more robust check the phase-shifted soliton solution can be recovered by taking ${\text{Re}[B]\rightarrow -\inf}$.  Applying the result of \ref{SGSolitonLimit} in the genus 1 case, the two solutions \refeq{SGEllipticV} in this limit become,
\begin{equation} \label{SGDefectSolitonLimit}
	e^{iv/2} \rightarrow \frac{1+i e^\Delta E}{1-i e^\Delta E}, \quad
	e^{iv/2} \rightarrow -\frac{1+i e^{-\Delta} E}{1-i e^{-\Delta} E}, \quad
	E=(-1)^\varepsilon \exp(\cosh\theta \,x - \sinh\theta\, t+x_0).
\end{equation}
The phase shift \refeq{jacobidefectphase} can be conveniently written as
\begin{equation} \label{jacobidefectphaseThetaEta}
	\frac{\jth{1}(\Delta)}{\jth{2}(\Delta)} = -i e^{\theta-\eta},
\end{equation}
where the rapidity $\theta=-\ln(p_1 q_1)/4$ and as before $\sigma=\exp(-\eta)$.
Then in the soliton limit \refeq{jacobidefectphaseThetaEta} becomes,
\begin{equation}
	-i e^{\theta-\eta} = \frac{\jth{1}(\Delta)}{\jth{2}(\Delta)}
	\rightarrow -i \tanh\left(\frac{\Delta}{2}\right),
\end{equation}
so that
\begin{equation}
	e^{\Delta} = \coth\left(\frac{\eta-\theta}{2}\right).
\end{equation}
Therefore the two finite-gap solutions in the soliton limit, \refeq{SGDefectSolitonLimit}, match the purely phase-shifted soliton solutions, \refeq{SGv1to1_cont} and \refeq{SGv1to1_discont}.

\p A feature of the genus 1 phase-shifted solutions, which does not have an analogue in the soliton case, is that for a given choice of distinct real branch points there exists a range of values for the defect parameter $\sigma$ such that the reality condition on $\Delta$,
\begin{equation} \label{realshift}
	\text{Im}[\Delta] = 0 \text{ or } \pi
\end{equation}
cannot be satisfied.
An example of this can be seen in \reffig{DeltaFig}(a).

\p An explanation for this gap comes from considering the bounds
\begin{align}
	-\frac{\jth{4}}{\jth{3}} \leq
	i\frac{\jth{1}(\Delta)}{\jth{2}(\Delta)} &
	\leq \frac{\jth{4}}{\jth{3}} &&\;\; \text{for Im}[\Delta]=0
	\label{0DeltaBound}
	\\
	i\frac{\jth{1}(\Delta)}{\jth{2}(\Delta)} \geq \frac{\jth{3}}{\jth{4}} \;\;\text{or}\;\; & i\frac{\jth{1}(\Delta)}{\jth{2}(\Delta)} \leq -\frac{\jth{3}}{\jth{4}} &&\;\; \text{for Im}[\Delta]=\pi
	\label{PiDeltaBound}
\end{align}
Using \refeq{jacobidefectphaseThetaEta} it can be seen that in the region
\begin{equation} \label{complexgap_thetaeta}
	\frac{\jth{4}}{\jth{3}} < e^{\theta-\eta} < \frac{\jth{3}}{\jth{4}}
\end{equation}
neither of the bounds for \refeq{0DeltaBound} and \refeq{PiDeltaBound} can be satisfied and hence there can be no real solutions of the form \refeq{defectjacobi}.

\p For complex conjugate branch points there is no gap, as demonstrated in \reffig{DeltaFig}(b).
The reason for this is that in this case $\jth{1}(\Delta)/\jth{2}(\Delta)$ is not bounded so there always exists a real solution to \refeq{jacobidefectphaseThetaEta} for any $\eta$.

\p It is interesting to note that the range \refeq{complexgap_thetaeta} can be  rewritten as
\begin{equation}\label{complexgap}
	\sqrt{-q_1} < \sigma < \sqrt{-p_1},
\end{equation}
which suggests that $-\sigma^2$ is a significant variable from the point of view of the Riemann surface. This is because if $-\sigma^2$ lies on the cut between $p_1$ and $q_1$ then the phase-shifted field $v$ to the right of the defect becomes complex.
The explanation for this is more apparent if the phase shift is rewritten in an explicit form as an integral of the holomorphic differential.
To accomplish this the ratio of theta functions is written as an elliptic function
\begin{equation}
	\frac{\theta_1(\Delta)}{\theta_2(\Delta)} = \frac{\jth{4}}{\jth{3}} \; \sc{\frac{\Delta \jth{3}^2}{2i}}.
\end{equation}
The expression for the inverse function given in \cite[\S 22.122]{Whittaker1962}
\begin{equation}
	\text{sc}^{-1}(z) = \int_0^z \frac{dt}{\sqrt{1+t^2}\sqrt{1+{q_1}t^2/{p_1}}}
\end{equation}
can, with the substitution $\lambda=-q_1 t^2$ be related to an integral of the holomorphic differential
\begin{equation}
	\text{sc}^{-1}(z) = \frac{\sqrt{-p_1}}{2C} \int_0^{-q_1 z^2} \omega, \quad
	q_1 z^2 \in \real,
\end{equation}
where the integral is performed on the upper sheet.
Then the phase shift for $\sigma>0$ up to periods of sc is written as
\begin{equation} \label{SGDeltaIntegralElliptic}
	\Delta = \int_0^{-\sigma^2} \omega
\end{equation}
where the integral is carried out over the upper sheet.

\p Now, in the case of real branch points $\int_0^{-\sigma^2}\omega$ is real if $-\sigma^2 \geq q_1$ and has imaginary part $i\pi$ if $-\sigma^2 \leq p_1$ but if $p_1 < -\sigma^2 < q_1$ then $\int_0^{-\sigma^2}\omega$ will be generically complex and the reality conditions for $\Delta$ will not be satisfied.
For conjugate branch points $\int_0^{-\sigma^2}\omega$ is real for $-\sigma^2 > (p_1+q_1)/2$ and has imaginary part $i\pi$ for $-\sigma^2 \leq (p_1+q_1)/2$ so $\Delta$ will satisfy the reality condition for any $\sigma$.

\section{Finite-gap solutions to the defect equations for sine-Gordon}
\label{sec:SGFiniteGapDefects}

\p For general genus, $g>1$, it seems to be very difficult to substitute directly the phase shift ansatz into the defect equations as was possible for $g=1$ because there does not appear to be a simple formula analogous to \refeq{derivformula} for the derivatives.
In any case, it would be interesting to find more general solutions involving soliton creation on a finite-gap background, analagous to the  previously known soliton solution discussed in \refsec{sec:SGSolCreation}, but this will require a different approach.

\p Following the discussion in \refsec{sec:DefectsAndBT}, a potentially useful idea is first to perform a B{\"a}cklund transformation on the finite-gap solution \refeq{SGfinitegap} for arbitrary genus. This will be achieved via a Darboux transformation of the Lax pair eigenfunctions corresponding to the finite-gap solutions. As before, some of the material that follows is review but included to make the paper more self-contained and to establish notation.

\subsection{A Darboux transformation for sine-Gordon}

With a change of coordinates
\begin{equation}
	\xi = \frac{t-x}{4}, \qquad
	\rho = \frac{t+x}{4}
\end{equation}
the sine-Gordon equation becomes
\begin{equation}
	u_{\xi\rho} = -4\sin u
\end{equation}
which is the compatibility condition of the Lax Pair,
\begin{equation} \label{SGLax}
	\psi_\rho = U(u) =
	\left( \begin{array}{ccc}
	0 & i \lambda e^{-iu} \\
	i e^{iu} & 0 \end{array} \right) \psi
	\qquad
	\psi_\xi = V(u) =
	\left( \begin{array}{ccc}
	-\frac{i u_\xi}{2} & i \\
	i \lambda^{-1} & \frac{i u_\xi}{2} \end{array} \right) \psi,
\end{equation}
similar to that given in \cite{Krichever1985}.

\p The next step is to find the appropriate form of the Darboux transformation which will connect the eigenfunction $\psi$ corresponding to the given $u$ with the Darboux transformed eigenfunction $\tilde\psi$ corresponding to the B{\"a}cklund transformed field $v$.
\begin{equation}
\begin{gathered}
	\psi_\rho = U(u)\psi,
	\qquad
	\tilde\psi_\rho = U(v)\tilde\psi,
	\\
	\psi_\xi = V(u)\psi,
	\qquad
	\tilde{\psi}_\xi = V(v)\tilde{\psi}.
\end{gathered}
\end{equation}
In this case, the question is to find a matrix $F(\rho,\xi,\lambda)$ such that,
\begin{subequations} \label{DarbouxEqs}
\begin{gather}
	\tilde\psi(\rho,\xi,\lambda) = F(\rho,\xi,\lambda) \psi(\rho,\xi,\lambda) \label{DarbouxPsi} \\
	F_\rho = U(v)\, F - F\, U(u) \\
	F_\xi = V(v)\, F - F\, V(u)
\end{gather}
\end{subequations}
is equivalent to the B{\"a}cklund transformation
\begin{subequations}
\begin{gather}
	\partial_\xi(v+u) = -\frac{4}{\sigma} \sin\left(\frac{v-u}{2}\right) \\
	\partial_\rho(v-u) = 4 \sigma \sin\left(\frac{v+u}{2}\right).
\end{gather}
\end{subequations}

\p Starting with the assumption that
\begin{equation}
	F(\rho,\xi,\lambda) = F_0(\rho,\xi) + \lambda F_1(\rho,\xi),
\end{equation}
the Darboux matrix is found to be
\begin{equation} \label{SGDarboux}
	F =
	\left( \begin{array}{ccc}
	i\lambda e^{\frac{i}{2}(u-v)} & \lambda \sigma \\
	\sigma & i\lambda e^{-\frac{i}{2}(u-v)} \end{array} \right)
\end{equation}
 and satisfies the above properties.
However, this form is not directly useful for generating solutions since the Darboux matrix has an explicit dependence on the unknown field $v$.  It is, therefore, useful to find a relationship between \refeq{SGDarboux} and the original field $u$.
To do this introduce a matrix of linearly independent solutions to \refeq{SGLax},
\begin{equation}
	\Psi(\lambda) =
	\left( \begin{array}{ccc}
	\Psi_{11}(\lambda) & \Psi_{12}(\lambda) \\
	\Psi_{21}(\lambda) & \Psi_{22}(\lambda) \end{array} \right).
\end{equation}
Following the argument of \cite{Gu1995}, since
\begin{equation}
	\Det{\tilde\Psi(\lambda)} = \Det{F(\lambda)} \Det{\Psi(\lambda)} = -\lambda(\lambda+\sigma^2) \Det{\Psi(\lambda)},
\end{equation}
the columns of $\tilde\Psi(\lambda)$ are linearly dependent for $\lambda=0$ or $-\sigma^2$ and therefore
\begin{equation} \label{DetPsi}
	\tilde\Psi(0)
	\left( \begin{array}{ccc}
	\alpha_1 \\
	\beta_1
	\end{array} \right) = 0'
	\qquad
	\tilde\Psi(-\sigma^2)
	\left( \begin{array}{ccc}
	\alpha_2 \\
	\beta_2
	\end{array} \right) = 0,
\end{equation}
for some constants $\alpha_i$, $\beta_i$, which are not both zero.
Assuming $\alpha_2 \neq 0$ let $b=\beta_2/\alpha_2$ and using $\tilde\Psi=F\Psi$ the following relations can be derived from \refeq{DetPsi},
\begin{subequations}
\begin{gather}
	\alpha_1 \Psi_{11}(0) + \beta_1 \Psi_{12}(0) = 0 \label{DetEq1} \\
	\frac{\Psi_{21}(-\sigma^2) + b \, \Psi_{22}(-\sigma^2)}{\Psi_{11}(-\sigma^2) + b \, \Psi_{12}(-\sigma^2)} = \frac{1}{i\sigma} e^{\frac{i}{2}(u-v)}. \label{DetEq2}
\end{gather}
\end{subequations}
The second equation \refeq{DetEq2} provides the needed explicit transformation $u \rightarrow v$,
\begin{equation} \label{SGDarbouxTrans}
	v = u + 2i \log\left[i\sigma \frac{\Psi_{21}(-\sigma^2) + b \, \Psi_{22}(-\sigma^2)}{\Psi_{11}(-\sigma^2) + b \, \Psi_{12}(-\sigma^2)} \right].
\end{equation}

\subsection{Multisoliton solutions from Darboux transformations}

As a test, equations \refeq{SGDarbouxTrans} can be used to derive the previously known one-to-two soliton solution to the defect equations discussed in \refeq{sec:SGSolCreation}.
Starting with the vacuum solution,
\begin{gather}
	u_0 = 0 \\
	\Psi_0 =
	\left( \begin{array}{ccc}
	\cos\left(\rho\sqrt{\lambda} + \frac{\xi}{\sqrt{\lambda}}\right) & i\sqrt{\lambda} \sin\left(\rho\sqrt{\lambda} + \frac{\xi}{\sqrt{\lambda}}\right) \\
	\frac{i}{\sqrt{\lambda}} \sin\left(\rho\sqrt{\lambda} + \frac{\xi}{\sqrt{\lambda}}\right) & \cos\left(\rho\sqrt{\lambda} + \frac{\xi}{\sqrt{\lambda}}\right)
	\end{array} \right),
\end{gather}
%

and using \refeq{SGDarbouxTrans} the result is
\begin{equation}
	u_1 = 2i \log\left[\frac{\sinh\left(\frac{\xi}{\sigma_1} - \rho\sigma_1\right) + i b_1 \sigma_1 \cosh\left(\frac{\xi}{\sigma_1} - \rho\sigma_1\right)}{\cosh\left(\frac{\xi}{\sigma_1} - \rho\sigma_1\right) + i b_1 \sigma_1 \sinh\left(\frac{\xi}{\sigma_1} - \rho\sigma_1\right)}\right],
\end{equation}
which after setting
\begin{equation}
	b_1 = \frac{i}{\sigma_1} \frac{1+ie^{x_{01}}}{1-ie^{x_{01}}}
\end{equation}
is simplified to
\begin{equation}
	u_1 = 2i \log\left[\frac{1-iE_1}{1+iE_1}\right],
\end{equation}
where
\begin{equation}
	E_i = \exp\left(2\sigma_i\rho - \frac{2\xi}{\sigma_i} - x_{0i}\right).
\end{equation}
With a change of coordinates back to $x,t$ this becomes the one soliton solution \refeq{SGsoliton} with the soliton rapidity $\theta_1$ given by $\sigma_1=\exp(-\theta_1)$.
The eigenfunction corresponding to $u_1$ is given by \refeq{DarbouxPsi} by
\begin{equation}
	\Psi_1 =
	\left( \begin{array}{ccc}
	\frac{i\lambda (1-i E_1)(1+E_\lambda^2) - \sigma_1 \sqrt{\lambda}(1+iE_1)(1-E_\lambda^2)}{2E_\lambda (1+iE_1)}
	&
	-\lambda \frac{i\sqrt{\lambda} (1-i E_1)(1-E_\lambda^2) - \sigma_1 (1+iE_1)(1+E_\lambda^2)}{2E_\lambda (1+iE_1)}
	\\
	-\frac{i\sqrt{\lambda}(1+iE_1)(1-E_\lambda^2) - \sigma_1 (1-iE_1)(1+E_\lambda^2)}{2 E_\lambda (1-iE_1)}
	&
	\frac{i\lambda (1+iE_1)(1+E_\lambda^2) - \sigma_1 \sqrt{\lambda} (1-i E_1)(1-E_\lambda^2)}{2(1-i E_1)E_\lambda}
	\end{array} \right),
\end{equation}
where
\begin{equation}
	E_\lambda = e^{i\left(\rho\sqrt{\lambda} + \xi/\sqrt{\lambda}\right)}.
\end{equation}
Using \refeq{SGDarbouxTrans} once more and letting
\begin{equation}
	b_2 = \frac{i}{\sigma_2} \frac{\varepsilon_2 + i\delta e^{x_{02}}}{\varepsilon_2 - i\delta e^{x_{02}}},
	\qquad \varepsilon_2 = \mp 1
\end{equation}
\begin{equation}
	\delta
	= \frac{\sigma_1+\sigma_2}{\sigma_1-\sigma_2}
	= \coth\left(\frac{\theta_2-\theta_1}{2}\right)§
\end{equation}
the expression for $u_2$ is
\begin{equation}
	u_2 = 2i\log\left[\frac{1 - i\delta E_1 + \varepsilon_2 i E_2 + \varepsilon_2 \delta^{-1} E_1 E_2 }{1 + i \delta E_1 - \varepsilon_2 i E_2 + \varepsilon_2 \delta^{-1} E_1 E_2 }\right].
\end{equation}
 This is precisely the two soliton solution \refeq{SGv1to2} appearing in the one-to-two soliton solution for the defect equations.

\subsection{Finite-gap eigenfunctions for sine-Gordon}
\label{sec:SGEigenfunctions}

\p Having established the effectiveness of this method for deriving solutions to the defect sewing equations in the purely solitonic case it is now time to turn to finite-gap solutions to the defect equations.

\p In order to find the Lax pair eigenfunctions that correspond to the finite-gap solutions \refeq{SGfinitegap} a construction similar to that given in \cite{Krichever1985} will be adopted.
Two solutions, $\psi^+_i$ and $\psi^-_i$, are sought having the following asymptotic form in the neighbourhood of $\lambda=0$,
\begin{equation}
	\begin{aligned}
	\psi_1^\pm(\xi,\rho,\lambda) &= e^{\pm i k_2\xi}                             \left(1 + \sum_{s=1}^{\infty} f_{s1}(\xi,\rho) \, k_2^{-s} \right) \\
	\psi_2^\pm(\xi,\rho,\lambda) &= \pm e^{\pm i k_2\xi} \, \frac{1}{\sqrt{\lambda}} \left(1 + \sum_{s=1}^{\infty} f_{s2}(\xi,\rho) \, k_2^{-s} \right)
	\end{aligned}
	\qquad{\rm where}\
	k_2 = \frac{1}{\sqrt{\lambda}},
\end{equation}
while in the neighbourhood of $\lambda=\infty$
\begin{equation}
	\begin{aligned}
	\psi_1^\pm(\xi,\rho,\lambda) &= e^{\pm i k_1\rho} \, c_1(\xi,\rho) 	                    \left(1 + \sum_{s=1}^{\infty} g_{s1}(\xi,\rho) \, k_1^{-s} \right) \\
	\psi_2^\pm(\xi,\rho,\lambda) &= \pm e^{\pm i k_1\rho} \, \frac{c_2(\xi,\rho)}{\sqrt{\lambda}} \left(1 + \sum_{s=1}^{\infty} g_{s2}(\xi,\rho) \, k_1^{-s} \right)
	\end{aligned}
	\qquad{\rm where}\
	k_1 = \sqrt{\lambda}.
\end{equation}
The Baker-Akhiezer functions matching these asymptotic forms are
\begin{subequations}
 \label{SGBA_old}
 \begin{align}
	\psi_1^\pm(\xi,\rho,\lambda) &= \frac{\theta(D) \, \theta\left(i B_1\rho + i B_2\xi + D \pm \int_0^{P^+}\omega\right)}{\theta\left(D \pm \int_0^{P^+}\omega\right) \, \theta(i B_1\rho + i B_2\xi + D)} \, e^{\pm i\Omega_1(P^+)\rho \pm i\Omega_2(P^+)\xi} \\
	\psi_2^\pm(\xi,\rho,\lambda) &= \pm \frac{\theta(D) \, \theta\left(i B_1\rho + i B_2\xi + B_3 + D \pm \int_0^{P^+}\omega\right)}{\theta\left(D \pm \int_0^{P^+}\omega\right) \, \theta(i B_1\rho + i B_2\xi + B_3 + D)} \, e^{\pm i\Omega_1(P^+)\rho \pm i\Omega_2(P^+)\xi + \Omega_3(\lambda)},
 \end{align}
\end{subequations}
for a constant vector $D$, and where the point $P^+=(+\mu, \lambda)$ lies on the upper sheet.
This expression also makes use of the integral,
\begin{equation}
	\Omega_3(\lambda) = -\frac{1}{2} \int_1^\lambda \frac{dt}{t} = -\frac{1}{2} \log(\lambda), \qquad
	B_{3i} = -\frac{1}{2} \oint_{b_i} \frac{d\lambda}{\lambda} = i\pi
\end{equation}
where the logarithm is taken to be principal valued so that $\Omega_3(1)=0$.

\p It is conceptually neater to think of these two solutions as corresponding to the two different points on the Riemann surface that have the same $\lambda$.  Since
\begin{equation}
	\int_0^{P^+} \omega = - \int_0^{P^-} \omega, \quad
	\Omega_1(P^+) = -\Omega_1(P^-), \quad
	\Omega_2(P^+) = -\Omega_2(P^-),
\end{equation}
 the two solutions \refeq{SGBA_old} may be rewritten as
\begin{subequations}
 \label{SGBA}
 \begin{align}
	\psi_1(\xi,\rho,P^\pm) &= \frac{\theta(D) \, \theta\left(i B_1\rho + i B_2\xi + D + \int_0^{P^\pm}\omega\right)}{\theta\left(D + \int_0^{P^\pm}\omega\right) \, \theta(i B_1\rho + i B_2\xi + D)} \, e^{i\Omega_1(P^\pm)\rho + i\Omega_2(P^\pm)\xi} \\
	\psi_2(\xi,\rho,P^\pm) &= \pm \frac{\theta(D) \, \theta\left(i B_1\rho + i B_2\xi + B_3 + D + \int_0^{P^\pm}\omega\right)}{\theta\left(D + \int_0^{P^\pm}\omega\right) \, \theta(i B_1\rho + i B_2\xi + B_3 + D)} \, e^{i\Omega_1(P^\pm)\rho + i\Omega_2(P^\pm)\xi + \Omega_3(\lambda)}.
 \end{align}
\end{subequations}
In addition, the overall factor of $\pm$ in $\psi_2$ can be viewed as a change of sheet for the logarithm where $\Omega_3(\lambda) \rightarrow \Omega_3(\lambda) + i\pi$.

\p To check that these eigenfunctions actually do correspond to the finite-gap solutions \refeq{SGfinitegap} it is key to note that
Baker-Akhiezer functions are uniquely defined up to a multiplicative function independent of $\lambda$ by their asymptotic forms at their singularities and the constant vector $D$ \cite{Krichever1985,AGANIE}.
The expansion of $\psi_{1\rho}$ and $\lambda\psi_2$ at $0$ and $\infty$ have the same form and therefore they are related by a function of $\xi,\rho$.  Comparing the coefficients of $\sqrt{\lambda}$ at $\infty$ it is found that $\psi_{1\rho}= i (c1/c2) \lambda\psi_2$.  Making similar comparisons one can verify that
\begin{equation} \label{SGLaxAsymptotics}
	\psi_\rho =
	\left( \begin{array}{ccc}
	0 & i\lambda \frac{c1}{c2} \\
	i\frac{c2}{c1} & 0 \end{array} \right) \psi
	\qquad
	\psi_\xi =
	\left( \begin{array}{ccc}
	\frac{c1_\xi}{c1} & i \\
	i \lambda^{-1} & \frac{c2_\xi}{c2} \end{array} \right) \psi,
\end{equation}
and to match with \refeq{SGLax} let
\begin{equation}
	c_1 = e^{-\frac{iu}{2}}, \qquad
	c_2 = e^{\frac{iu}{2}}.
\end{equation}
Evaluating \refeq{SGBA} at $P=\infty$ and noting that $\int_0^{\infty} \omega = -i\pi$ then
\begin{equation}
	\frac{c_2}{c_1} = e^{iu} = \frac{[\theta(i B_1 \rho + i B_2 \xi + D)]^2}{[\theta(i B_1 \rho + i B_2 \xi + D + i\pi)]^2},
\end{equation}
which, after a change of variables back to $x,t$, gives \refeq{SGfinitegap}.

\subsection{Finite-gap Darboux transformation}

\p Now, returning to the original problem, suppose there is  a defect at $x=x_D$ with parameter $\sigma>0$ and  that for $x<x_D$ the field, $u$, is a finite-gap solution \refeq{SGfinitegap} for any genus.
Then, applying the transformation \refeq{SGDarbouxTrans} with
\begin{equation}
	\Psi(\lambda)
	=
	\left( \begin{array}{ccc}
	\psi_1(P^+) & \psi_1(P^-) \\
	\psi_2(P^+) & \psi_2(P^-) \end{array} \right)
	\qquad P^\pm = (\pm\mu,\lambda),
\end{equation}
 the corresponding field, $v$, in the region $x>x_D$ is
\begin{equation} \label{SGDarbouxTransformedFiniteGap}
\begin{gathered}
	v = 2i \log \left[\frac{\theta\left(z+i\pi+\int_0^{-\sigma^2}\omega\right) - K E_\sigma \theta\left(z+i\pi-\int_0^{-\sigma^2}\omega\right)}{\theta\left(z+\int_0^{-\sigma^2}\omega\right) + K E_\sigma \theta\left(z-\int_0^{-\sigma^2}\omega\right)} \right] \\
	z = i B_1\rho + i B_2\xi + D, \quad
	K = b \; \frac{\theta\left(D+\int_0^{-\sigma^2}\omega\right)}{\theta\left(D-\int_0^{-\sigma^2}\omega\right)} \\
	E_\sigma = \exp\left[-2i(\Omega_1(-\sigma^2)\rho + \Omega_2(-\sigma^2)\xi)\right],
\end{gathered}
\end{equation}
where $-\sigma^2$ is a point on the upper sheet for the purposes of integration.

\subsection{Reality conditions}

For $v$ to be real there must be some constraints on $-\sigma^2$ and $b$.
For the chosen basis of cycles $a_i$ and $b_i$ it can be shown \cite{AGANIE} that
\begin{equation}\label{SGConjugate}
	\overline{C_{ij}} = -C_{ij}, \qquad
	\overline{\theta(r,B)} = \theta(\bar r + i\pi\kappa + i\pi, B),
\end{equation}
where the bar denotes complex conjugation and $\kappa$ is given by \refeq{SGfinitegap}.
It then follows from \refeq{SGSecondB} that
\begin{equation}
	\overline{B_1} = -B_1, \qquad
	\overline{B_2} = -B_2, \qquad
\end{equation}
The form of $D$ is already restricted by the reality conditions for $u$ and has the property
\begin{equation}\label{SGConjugateD}
	\bar D = D -i\pi \kappa - 2i\pi \varepsilon
\end{equation}

\p If $\phi$ is the argument of the logarithm in \refeq{SGDarbouxTransformedFiniteGap} then the reality of $v$ is equivalent to $\phi\bar\phi = 1$.  The complex conjugate of $\phi$ is
\begin{equation*}
	\bar\phi =
	\frac{\theta\left(z+\overline{\int_0^{-\sigma^2}\omega}\right) -
	\bar K \bar E_\sigma \theta\left(z-\overline{\int_0^{-\sigma^2}\omega}\right)}
	{\theta\left(z+i\pi+\overline{\int_0^{-\sigma^2}\omega}\right) +
	\bar K \bar E_\sigma \theta\left(z+i\pi-\overline{\int_0^{-\sigma^2}\omega}\right)},
\end{equation*}
so in order for $v$ to be real each element of $\int_0^{-\sigma^2}\omega_i$ is required either to be real or to have imaginary part $i\pi$.
The holomorphic differential $\omega$ is imaginary in $[p_i,q_i]$ and real for $\lambda\in\real$ but not on a cut.
Since, from \refeq{PeriodsAsLineIntegrals}, $\int_{p_i}^{q_i} \omega_j = i\pi \delta_{ij}$ it is required that $-\sigma^2$ does not lie in any of the intervals $[p_i,q_i]$.

\p Turning to $E_\sigma$, the Abelian differentials of the second kind $d\Omega_1$, $d\Omega_2$ are real when evaluated on a cut $[p_i,q_i]$ and imaginary for $\lambda\in\real$ but not on a cut.
The a-periods of $d\Omega_1$ and $d\Omega_2$ are normalised to zero so by \refeq{PeriodsAsLineIntegrals} $\int_{p_i}^{q_i} d\Omega_1 = \int_{p_i}^{q_i} d\Omega_2 = 0$.
Therefore the restriction that $-\sigma^2$ does not lie in any $[p_i,q_i]$ guarantees that $\Omega_1(-\sigma^2)$ and $\Omega_2(-\sigma^2)$ are purely imaginary and hence $\bar E_\sigma = E_\sigma$.
Finally,  $b$ is required to be such that $\bar K = -K$.

\subsection{Phase-shifted and soliton limits}

Based on the results in \cite{Matveev1991} for a Darboux transformation on an arbitrary background for the KdV and nonlinear Schr\"{o}dinger equations, it is expected that the expression \refeq{SGDarbouxTransformedFiniteGap} comprises a soliton on the background of the original finite-gap solution. A recent example of constructing solutions containing KdV solitons on an elliptic (cnoidal) background is provided in \cite{Arancibia2015}.
In appendix \ref{sec:partialSolitonLimit} it is found that this is the case by demonstrating that \refeq{SGDarbouxTransformedFiniteGap} has the same form as an expression obtained by taking the limit of a genus $g+1$  solution to sine-Gordon in which the points in one of the pairs of branch points coalesce at the point $-\sigma^2$.

\p Just as was the case for the one-to-two soliton solution, it is possible to obtain purely phase-shifted solutions from \refeq{SGDarbouxTransformedFiniteGap} by taking the limits $b\rightarrow 0$ or $b\rightarrow\infty$.
Thus, respectively,
\begin{equation}
	e^{iv/2} \rightarrow \frac{\theta\left(z+\int_0^{-\sigma^2}\omega\right)}{\theta\left(z+i\pi+\int_0^{-\sigma^2}
\omega\right)}, \qquad
	e^{iv/2} \rightarrow -\frac{\theta\left(z-\int_0^{-\sigma^2}\omega\right)}{\theta\left(z+i\pi-\int_0^{-\sigma^2}
\omega\right)},
\end{equation}
with the phase shifts for each case being
\begin{equation}
	\Delta = \int_0^{-\sigma^2} \omega, \qquad
	\Delta = Bn-\int_0^{-\sigma^2} \omega,
\end{equation}
for any $n\in\integers^g$ where the sum of $n_i$ is odd.
This is therefore the natural generalisation to higher genera of the integral expression for $\Delta$ \refeq{SGDeltaIntegralElliptic} obtained in the genus 1 case.

\p The one-to-two soliton solution discussed in \refsec{sec:SGSolCreation} can also be recovered from the genus 1 finite-gap solutions to the defect equations by taking the soliton limit in which $q_1 \rightarrow p_1 =: \lambda_1$.
The details of this limit for a  genus $g$ solution to sine-Gordon are repeated for convenience in \ref{sec:FullSolitonLimit}.

\p For the finite-gap field to the left of the defect \refeq{SGfinitegap} it is useful first to parameterise $D$ by
\begin{equation}
	D = -x_\theta + i\pi/2 + i\pi \varepsilon - B/2, \quad x_\theta \in \real
\end{equation}
so that in the soliton limit \refeq{SGfinitegap} becomes
\begin{equation} \label{SG1SolitonLimit}
	e^{i u/2} \rightarrow \frac{1+iE_\theta}{1-iE_\theta}, \quad
	E_\theta = (-1)^{\varepsilon} \exp\left[ \cosh(\theta)x - \sinh(\theta)t - x_\theta \right],
\end{equation}
where
\begin{equation}
	\theta = -\frac{1}{2} \log(-\lambda_1).
\end{equation}
 Setting $\varepsilon = 0$ \refeq{SG1SolitonLimit} matches the one soliton solution \refeq{SGsoliton}.

\p It was  seen in \refsec{sec:SG1GensuPhaseShifted} that in the soliton limit
\begin{equation}
	\exp\left[ \int_0^{-\sigma^2} \omega \right] \rightarrow \delta = \coth\left(\frac{\eta-\theta}{2}\right)
\end{equation}
where $\sigma = \exp(-\eta)$, and this can be confirmed by directly integrating the holomorphic differential in the soliton limit \refeq{holomorphicSoliton} for the case where $\lambda_0=0$ and $g=1$.
From the soliton limit of the Abelian differentials of the second kind \refeq{SGSecondKindDiff_SolitonLimit} the corresponding integrals are now, assuming $\sigma>0$
\begin{equation}
	\Omega_1(-\sigma^2) \rightarrow \int_0^{-\sigma^2} \frac{1}{2\sqrt{\lambda}} = i\sigma, \qquad
	\Omega_2(-\sigma^2) \rightarrow \int_\inf^{-\sigma^2} -\frac{1}{2\lambda\sqrt{\lambda}} = \frac{1}{i\sigma}
\end{equation}
where the path of integration for $\Omega_2$ avoids the singularity at $0$.

\p For the field to the right of the defect \refeq{SGDarbouxTransformedFiniteGap} set $K = \pm i\exp(-x_\eta)$ for $x_\eta \in\real$ so that in the soliton limit
\begin{equation}
	K E_\sigma \rightarrow \pm i E_\eta = \pm i\exp\left[\cosh\eta\,x - \sinh\eta\,t - x_\eta \right]
\end{equation}
expressed in the original $x,t$ coordinates.
Finally, in the soliton limit \refeq{SGDarbouxTransformedFiniteGap} becomes
\begin{equation}
\begin{gathered}
	e^{iv/2} \rightarrow \frac{(1+i\delta E_\theta) \pm iE_\eta (1+i\delta^{-1} E_\theta)}{(1-i\delta E_\theta) \mp iE_\eta (1-i\delta^{-1} E_\theta)} \\
\end{gathered}
\end{equation}
which is precisely the two soliton field \refeq{SGv1to2} to the right of the defect in the one-to-two soliton solution discussed in \refsec{sec:SGSolCreation}.
 The conclusion is that the pair of fields $u$, \refeq{SGfinitegap}, and $v$, \refeq{SGDarbouxTransformedFiniteGap} are an algebro-geometric generalisation of the known one-to-two soliton solution for the sine-Gordon type I defect in which a soliton is created at an undetermined time by the defect on a finite-gap background.

\section{Soliton solutions to the defect equations for KdV}

For the KdV equation,
\begin{equation}\label{KdV}
	u_t-6uu_x+u_{xxx}=0
\end{equation}
the type I integrable defect placed at $x=x_D$ can be written in terms of the potentials $p,q$ where $p_x=u$ and $q_x=v$ are the fields in the regions $x<x_D$ and $x>x_D$ respectively.  The defect conditions at the point $x=x_D$ are then \cite{Corrigan2006}.
\begin{subequations}
 \label{KdVdefectEq}
 \begin{align}
  p_x + q_x & = 2\sigma + \frac{1}{2} (p-q)^2 \label{defectEq1} \\
  p_t + q_t & = 2(p_x^2 + p_x q_x + q_x^2) - (p-q)(p_{xx} - q_{xx})\label{defectEq2} \\
  p_{xx}+q_{xx}&=(p-q)(p_x-q_x),
 \end{align}
\end{subequations}
where $\sigma\in\real$ is again the defect parameter.
Note, it is necessary  to specify that $p_{xx}+q_{xx}=(p-q)(p_x-q_x)$ explicitly as one of the sewing conditions.  On the full line the third condition would be the $x$ derivative of the first \refeq{defectEq1} but, because the defect conditions are restricted to the point $x=x_D$, the space derivatives are frozen.
Before repeating the arguments used for sine-Gordon to derive finite-gap solutions to the type I defect equations for KdV it is worth first reviewing the soliton solutions.

\subsection{Purely phase-shifted soliton solutions}

A single soliton solution to \refeq{KdV} can be described by,
\begin{equation}\label{pSoliton}
	p=p_0 - \frac{2aE}{1+E} + c_0 x + 3c_0^2 t, \qquad E=\exp\left[a\left(x-\left(a^2 - 6 c_0\right) t - x_0 \right)\right],
\end{equation}
which in terms of the original field becomes,
\begin{equation} \label{KdV1Soliton}
	u = p_x
	= -\frac{2a^2 E}{(1+ E)^2} + c_0
	= -\frac{a^2}{2} \text{sech}^2\left[\frac{a}{2} \left(x-\left(a^2 - 6c_0\right) t - x_0 \right)\right] + c_0,
\end{equation}
where $p_0, c_0, x_0, a$ are real constants.

\p In \cite{Corrigan2006} purely phase-shifted soliton solutions to the defect equations were found but only solitons with $c_0=0$ were examined.
Soliton solutions with arbitrary $c_0$ can be found by taking $p$ to correspond to the single soliton solution above and $q$ to be a phase-shifted single soliton
\begin{align}
	q &= q_0 - \frac{2a\shift E}{1+\shift E} + c_0 x + 3c_0^2 t \label{qSoliton} \\
	v &= q_x = -\frac{2a^2 \shift E}{(1+\shift E)^2} + c_0. \label{vSoliton}
\end{align}
Then introducing $\chi$ such that $\sigma=-\chi^2/4 + c_0$ it is found that \refeq{defectEq1} implies
\begin{equation} \label{KdVSolitonPhase}
	\chi^2=(p_0-q_0)^2, \qquad \shift = \frac{p_0-q_0 - a}{p_0-q_0 + a} = \frac{\abs{\chi} \mp a}{\abs{\chi} \pm a}.
\end{equation}
It can then be checked that the second defect equation, eq\refeq{defectEq2}, is identically true.

\p The phase shift in \refeq{KdVSolitonPhase} is given by $(\abs{\chi}-a)/(\abs{\chi}+a)$ in \cite{Corrigan2006} since there the positive square root for $p_0 - q_0$ was chosen. However,  the negative choice appears to be valid from the point of view of the defect equations since only $(p_0-q_0)^2$ is fixed.
An alternative source of the same sign ambiguity comes from noting that $a\rightarrow-a$ leaves the original field \refeq{KdV1Soliton} invariant.

\p As noted in \cite{Corrigan2006}, the phase shift $\Delta$ has some interesting features.
If $\shift$ is negative then the denominator of \refeq{qSoliton} will be zero for some value of $x,t$ and the solution $v=q_x$ will have a singularity.
If $a=\abs{\chi}$ then $\shift=0$ or $\shift\rightarrow\infty$ and, in either case, $v=q_x=c_0$ so the soliton is destroyed by the defect.

\p Just as was the case with sine-Gordon, the ability of a defect to destroy a soliton coupled with the fact that the defect equations \refeq{KdVdefectEq} are invariant under an exchange of $p$ and $q$ implies that there exists a solution where a soliton is created.
Such a one-to-two soliton solution was considered in \cite{Corrigan2006} for $c_0=0$.
Here, the method of Darboux transformation will be employed, as in the sine-Gordon case, to find the one-to-two soliton solution for the case of arbitrary $c_0$ and examine how energy is conserved in the soliton creation process for $c_0=0$.

\subsection{Lax pair and Darboux transformation}

The KdV equation \refeq{KdV} is the compatibility condition of the Lax pair,
\begin{subequations}
 \label{KdVLax}
 \begin{align}
  \psi_{xx} & = (u-\lambda) \psi \label{KdVLax1} \\
  \psi_t  & = (2u + 4\lambda) \psi_x - u_x \psi \label{KdVLax2},
 \end{align}
\end{subequations}
with eigenfunction $\psi(x,t,\lambda)$ and spectral parameter $\lambda\in\complex$.

\p The well known \cite{Matveev1991,Gu2005} Darboux transformation for this Lax pair is constructed as follows.
If $u_0$ and $\psi_0$ satisfy \refeq{KdVLax} then so too does
\begin{equation} \label{KdVdarboux}
	u_1=u_0 - 2\partial_{xx}\ln(f_1), \qquad \psi_1 = \psi_{0x} - \psi_0 \, \partial_{x}\ln(f_1)
\end{equation}
for ${f_1(x,t)=\psi_0(x,t,\sigma_1)}$ with some choice of $\sigma_1$.

\p The potentials $p,q$ such that $p_x=u_0$ and $q_x=u_1$ also satisfy a B{\"a}cklund transformation of the same form as the defect equations \refeq{KdVdefectEq} but applied to all $x$, provided that
\begin{equation} \label{potentialKdV}
	p_t - 3p_x^2 + p_{xxx} = 0
\end{equation}
To show this let $p_x=u_0$ and $q_x=u_1$ and integrate $p_x-q_x=2\partial_{xx}\ln(f)$ once with respect to $x$.  Since a function of $t$ can be absorbed into $q$ without changing $u_1$, let
\begin{equation} \label{pq}
	p-q=2\partial_{x}\ln(f_1).
\end{equation}
Then the expression for $u_1$ in \refeq{KdVdarboux} can, after using (6.8a) for $f_1$, be seen to be
\begin{equation} \label{backlund1}
	p_x + q_x = 2\sigma_1 + \frac{1}{2} (p-q)^2.
\end{equation}
One can also confirm that
\begin{equation} \label{backlund2}
	p_t + q_t = 2(p_x^2 + p_x q_x + q_x^2) - (p-q)(p_{xx} - q_{xx})
\end{equation}
holds for all $x,t$ by using \refeq{pq} and its derivatives to eliminate $q$ from \refeq{backlund2} and \refeq{KdVLax} to eliminate $f_1$ leaving the constraint \refeq{potentialKdV}, which is assumed.
Equations \refeq{backlund1} and \refeq{backlund2} then constitute a B{\"a}cklund transformation from the field $u_0=p_{x}$ to $u_1=q_{x}$, which is satisfied for all $x,t$.

\p Note that for a given $u_0$ is always possible to choose $p$ such that \refeq{potentialKdV} is true.
By inserting $u_0=p_x$ into the KdV equation \refeq{KdV} and integrating once with respect to $x$ one finds
\begin{equation}
	\frac{\partial}{\partial x} \left[p_t - 3p_x^2 + p_{xxx}\right] = 0
\end{equation}
and a function of $t$ can be absorbed into $p$ without changing $u_0$.

\p In addition, if $p$ satisfies \refeq{potentialKdV} then so too does $q$ since differentiating \refeq{backlund1} twice to obtain an expression for $(p-q)(p_{xx}-q_{xx})$ and substituting this into \refeq{backlund2} gives
\begin{equation}
	p_t + q_t = 3p_x^2 + 3q_x^2 - p_{xxx} - q_{xxx}.
\end{equation}
Therefore, if the Darboux transformation is repeatedly applied then each potential and its successive Darboux transformed potential will satisfy the B{\"a}cklund equations \refeq{backlund1} and \refeq{backlund2}.

\subsection{Soliton creation}
\label{sec:KdVsolitoncreation}

Starting from the solution,
\begin{equation}
	p=p_0 - a + c_0 x + 3c_0^2 t, \qquad
	u_0=p_x=c_0,
\end{equation}
first solve the Lax pair to find the corresponding eigenfunction,
\begin{equation}
	\psi_0(x,t,\lambda;b) = E^{-1/2} + b E^{1/2},
\end{equation}
for some constant $b$, and where
\begin{equation}
	E=\exp\left[\varphi (x-(\varphi^2-6c_0)t)\right], \quad
	\lambda=-\frac{\varphi^2}{4} + c_0.
\end{equation}
Note that the chosen $p$ satisfies the constraint \refeq{potentialKdV} so the potentials $p$ and $q_1$ are related by the B{\"a}cklund transformation over all $x,t$.

\p Performing the Darboux transformation \refeq{KdVdarboux} with
\begin{equation}
	f_1(x,t)=\psi_0(x,t,\lambda_1;b_1), \quad
	\lambda_1=-\frac{a^2}{4}+c_0,
\end{equation}
for some choice of $b_1$, gives,
\begin{subequations}
\begin{align}
q_1 &= p_0 - \frac{2a b_1 E_1}{1+b_1 E_1} + c_0 x + 3c_0^2 t , \label{KdVq1} \\
\psi_1(x,t,\lambda;b) &= \frac{(bE-1)(b_1E_1+1)\varphi - (b_1E_1-1)(bE+1)a}{2E^{1/2} (1+b_1E_1)},
\end{align}
\end{subequations}
where $E_1$ is defined by
\begin{equation}
E_1=\exp\left[a(x-(a^2-6c_0) t)\right]
\end{equation}
If $b_1\geq0$ then \refeq{KdVq1} is the one soliton potential \refeq{pSoliton}.

\p Applying a further Darboux transformation with
\begin{equation}
	f_2(x,t)=\psi_0(x,t,\lambda_2;b_2), \quad
	\lambda_2=-\frac{\chi^2}{4}+c_0,
\end{equation}
leads to the two soliton potential
\begin{equation}
q_2 = p_0 - a + \frac{(a^2-\chi^2)(1+b_1 E_1 + b_2 E_2 + b_1 b_2 E_1 E_2)}{(a-\chi)(1-b_1 b_2 E_1 E_2)-(a+\chi)(b_1 E_1 - b_2 E_2)} + c_0 x + 3c_0^2 t,
\end{equation}
where
\begin{equation}
E_2=\exp\left[\chi(x-(\chi^2-6c_0) t)\right].
\end{equation}
For $q_2$ to be regular either $b_1\leq0$, $b_2\geq0$ and ${\abs{a}>\abs{\chi}}$ or $b_1\geq0$, $b_2\leq0$ and $\abs{a} < \abs{\chi}$.
If $\abs{a} = \abs{\chi}$ then the Darboux transformation eliminates the soliton created in the first transformation.

\p The situation of interest is the case where the fields on both sides of the defect are regular. So it will be assumed from now on that $b_1\geq0$, $b_2\leq0$ and $\abs{a} < \abs{\chi}$.
 Thus, returning to the original problem where a single soliton is incident on a defect at $x=x_D$ with parameter $\sigma=-\chi^2/4 + c_0$, there is a solution where the potential for the field is $p$ for $x<x_D$ and $q$ for $x>x_D$,
\begin{subequations}
 \label{1to2Defect}
 \begin{align}
 	p &= p_0 - \frac{2a E_1}{1+E_1} + c_0 x + 3c_0^2 t \label{p1to2Defect} \\
	q &= p_0 - a + \frac{(a^2-\chi^2)(1+E_1-E_2-E_1E_2)}{(a-\chi)(1+E_1E_2)-(a+\chi)(E_1+E_2)} + c_0 x + 3c_0^2 t, \label{q1to2Defect}
 \end{align}
\end{subequations}
\begin{equation*}
	E_1=\exp\left[a\left(x - (a^2-6c_0)t - x_a \right)\right], \quad
	E_2=\exp\left[\chi\left(x - (\chi^2-6c_0)t - x_\chi \right)\right].
\end{equation*}
As was the case for sine-Gordon the potential for the purely phase-shifted solutions \refeq{qSoliton} can be recovered from \refeq{q1to2Defect} in the limits $x_{\chi}\rightarrow\pm\infty$.

\p The defect has a conserved energy and momentum by construction \cite{Corrigan2006} but as a check the total energy an infinite time before and after the original soliton passes through the defect and the additional soliton is created can be calculated, assuming the additional soliton is created at a finite time.
The conserved energy is the energy in the fields
\begin{equation}
	E_p + E_q = \int_{-\infty}^0 \left[(p_x)^3 + \frac{1}{2} (p_{xx})^2 \right]dx +
	\int_0^\infty \left[(q_x)^3 + \frac{1}{2} (q_{xx})^2 \right]dx
\end{equation}
plus the defect contribution \cite{Corrigan2006} \footnote{Eq(9.25) in \cite{Corrigan2006} has a sign error which is corrected in eq\refeq{ED}}
\begin{equation} \label{ED}
	E_D = -(p-q)\left.\left[(p_x^2 + p_x q_x + q_x^2) - (p-q)^2 \left(\sigma + \frac{3}{20} (p-q)^2\right)\right]\right|_{x=x_D}.
\end{equation}
To avoid issues with infinite energy assume for the moment that $c_0=0$.
Then the energy of a single KdV soliton on the full line is $-\abs{a}^5/5$.  For pure soliton solutions the energy is additive so the energy of the two soliton solution on the full line is $-(\abs{a}^5+\abs{\chi}^5)/5$.
In the initial configuration, with $t\rightarrow-\infty$ and $p$, $q$ defined by \refeq{1to2Defect} and $c_0=0$, $p-q \rightarrow \abs{\chi}$ so $E_D \rightarrow -\abs{\chi}^5/10$ (using $\sigma = -\chi^2/4$).
In the final configuration, $t\rightarrow\infty$,  $p-q \rightarrow -\abs{\chi}$ and $E_D \rightarrow \abs{\chi}^5/10$.
The change in the defect energy, $\abs{\chi}^5/5$, then precisely compensates for the energy of the additional soliton $-\abs{\chi}^5/5$.

\p The one-to-two soliton solution \refeq{1to2Defect} is a family of solutions parameterised by the initial position of the created soliton $x_\chi$, however, it is possible to pick out particular solutions with additional constraints.

\p For example, it follows from the first defect equation \refeq{defectEq1} that at infinite times $p-q=\pm\abs{\chi}$ at $x=x_D$ so imposing initially that $p-q=-\abs{\chi}$ at $(x,t)= (x_D, -\infty)$ the defect energy cannot be greater as $t\rightarrow\infty$ and will be unable to compensate for the (negative) energy of an additional soliton produced at a finite time.
This additional constraint picks out the purely phase-shifted solution \refeq{qSoliton} with $p_0-q_0=-\abs{\chi}$.

\section{Finite-gap solutions for KdV with type I defects}
\subsection{Finite-gap solutions on the full line}

\begin{figure}
	\center
	\includegraphics[width=15cm]{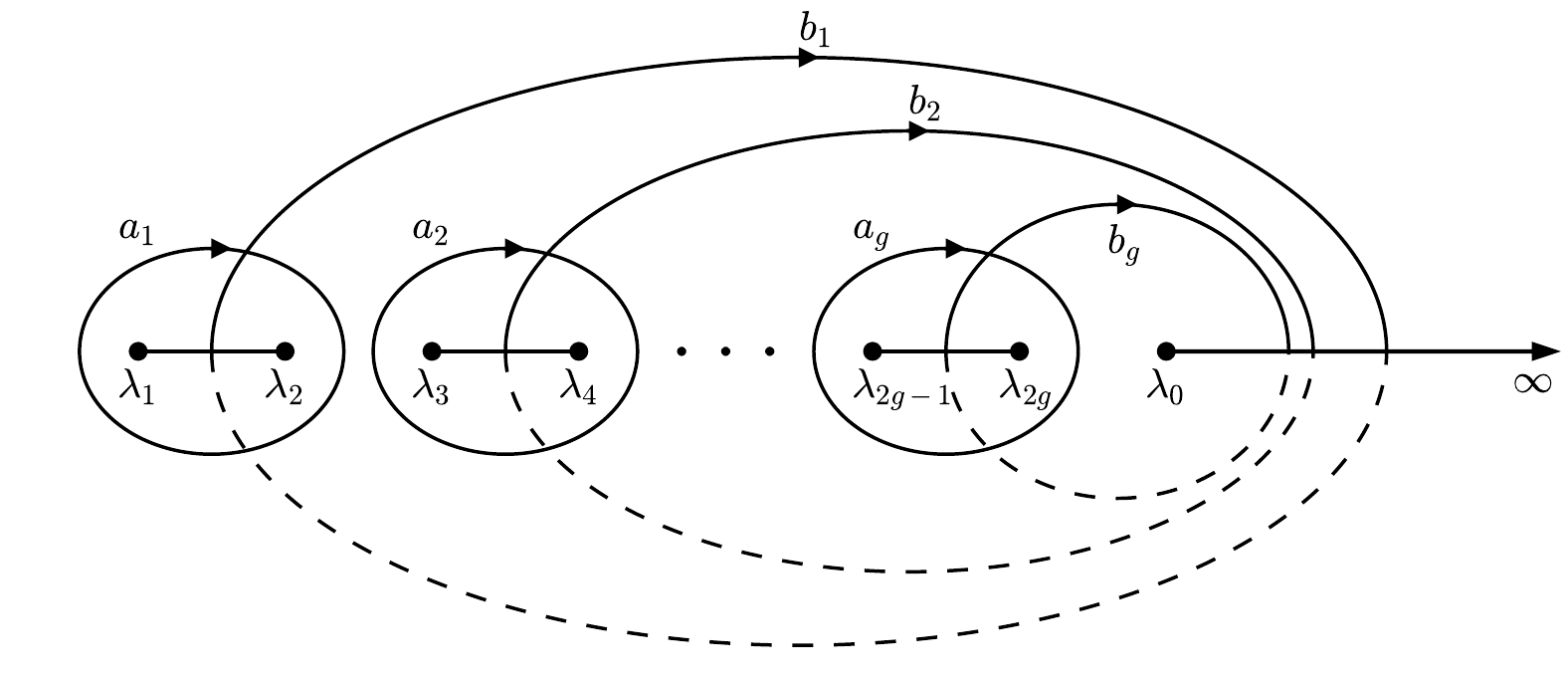}
	\caption{The branch points at $\lambda_i$ and $\infty$ and the branch cuts between them represented by solid lines.
	The basis of cycles are labeled $a_i$, $b_i$ and the parts of the cycles on the upper sheet have solid lines while the dashed lines are on the lower sheet.}
	\label{homology}
\end{figure}

The finite-gap solutions for KdV are characterised by a choice of branch points on the hyperelliptic Riemann surface \cite{AGANIE}
\begin{equation} \label{curve}
	\mu^2= (\lambda-\lambda_0) \prod\limits_{i=1}^{2g}(\lambda-\lambda_i), \quad
	\lambda \in \real, \quad
	\lambda_1 < \lambda_2 < \dots < \lambda_{2g} < \lambda_0.
\end{equation}
This surface has branch points at $\lambda=\lambda_i$ and at $\lambda=\infty$, and the branch cuts are taken to be in the intervals $[\lambda_1,\lambda_2], \dots,[\lambda_{2g-1},\lambda_{2g}], [\lambda_0,\infty]$ as shown in \reffig{homology}.
The upper sheet is defined analogously to the definition for sine-Gordon provided in \refsec{sec:SGFiniteGap}, similarly introducing the basis of cycles, $a_i$, $b_i$ shown in \reffig{homology}.

\p Solutions to the KdV equation \refeq{KdV} are then given by,
\begin{equation} \label{AGKdVSol}
	u(x,t) = -2 \partial_{xx} \log\left[\theta(Ux-4Wt+D, B)\right] - 2c_1.
\end{equation}

\p The normalized holomorphic differentials $\omega$ and Riemann matrix $B$ are defined in the same way as they were for sine-Gordon by \refeq{holomorphicDiff}, \refeq{holomorphicNormalization} and \refeq{RiemannMatrix}.
For KdV the abelian integrals of the second kind \cite{AGANIE}
\begin{equation}
	\Omega_1(P) = \int_{\lambda_{0}}^P d\Omega_1(P), \qquad
	\Omega_3(P) = \int_{\lambda_{0}}^P d\Omega_3(P),
\end{equation}
have the asymptotic forms
\begin{subequations}
 \label{KdVSecondKindExpansion}
 \begin{align}
	\Omega_1 & \rightarrow k - \frac{c_1}{k} + O(k^{-2}) , & \lambda \rightarrow \infty \label{omega1} \\
	\Omega_3 & \rightarrow k^3 - \frac{c_3}{k} + O(k^{-2}) , & \lambda \rightarrow \infty,
 \end{align}
\end{subequations}
with local parameter $k=i\sqrt{\lambda}$.
The corresponding differentials are
\begin{subequations}
 \label{KdVSecondKindDiff}
 \begin{align}
	d\Omega_1 &= \frac{i\lambda^g}{2\mu} d\lambda + \sum_{i=1}^g \alpha_i \omega_i, \label{KdVomega1Diff} \\
	d\Omega_3 &= -\frac{3i}{4} \left( \frac{2\lambda^{g+1} - \widehat{E} \lambda^g}{\mu} \right) d\lambda + \sum_{i=1}^g \beta_i \omega_i, \qquad \widehat{E} = \sum_{i=0}^{2g} \lambda_i,
 \end{align}
\end{subequations}
where the constants $\alpha_i, \beta_i$ are fixed by the normalisation conditions
\begin{equation} \label{KdVsecondNormalization}
	\oint_{a_i} d\Omega_1 = 0, \qquad
	\oint_{a_i} d\Omega_3 = 0,
\end{equation}
and the $g$ dimensional vectors $U,W$ are defined to be
\begin{equation}
	U_i = \oint_{b_i} d\Omega_1, \qquad
	W_i = \oint_{b_i} d\Omega_3.
\end{equation}

\p Again, $U$ and $W$ can be conveniently written in terms of the holomorphic differential normalization constants $C_{ij}$ using the Riemann bilinear relation \refeq{RiemannBilinear}.
In this case,
\begin{align}
	d\Omega_1 &= [z^{-2} + O(1)]dz, \quad z=i\lambda^{-1/2} \\
	d\Omega_3 &= [z^{-4} + O(1)]dz, \quad z=\frac{i}{3^{1/3}}\lambda^{-1/2},
\end{align}
and using the normalization conditions \refeq{holomorphicNormalization} and \refeq{KdVsecondNormalization} it is found that
\begin{align}
	U_i &= 2i C_{ig} \label{UFormula} \\
	W_i &= -2i \left(C_{i(g-1)} + \frac{1}{2} C_{ig} \sum_{k=0}^{2g} \lambda_k \right). \label{WFormula}
\end{align}

\p By directly expanding \refeq{KdVSecondKindDiff} and comparing with \refeq{KdVSecondKindExpansion} it follows that
\begin{equation} \label{c1}
	c_1 = 2i \sum_{k=1}^g \alpha_k C_{kg} -\frac{1}{2} \sum_{i=0}^{2g} \lambda_i
\end{equation}
\begin{equation} \label{c3}
	c_3 = 2i \sum_{k=1}^g \beta_k C_{kg} + \frac{3}{4} \sum_{i=0}^{2g} \lambda_i^2 - \frac{3}{8}\left(\sum_{i=0}^{2g} \lambda_i\right)^2.
\end{equation}

\subsection{Finite-gap Darboux transformation}

Using the same argument as was used for sine-Gordon finite-gap solutions to the defect equations will be derived by performing a Darboux transformation which corresponds to the defect sewing equations applied to the whole line.

\p Using similar reasoning to that contained in \refsec{sec:SGEigenfunctions}, it can be shown \cite{AGANIE} that the basis of eigenfunctions for the Lax pair \refeq{KdVLax}, which corresponds to the finite-gap solutions \refeq{AGKdVSol}, is given by,
\begin{equation} \label{KdVBA}
	\psi(x,t,P) = \frac{\theta\left(\int_\infty^P \omega + Ux-4Wt+D\right) \theta(D)}{\theta\left(\int_\infty^{P}\omega + D\right) \theta(Ux-4Wt+D)} e^{x\Omega_1(P) - 4t\Omega_3(P)}.
\end{equation}
The initial eigenfunction for the Darboux transformation is taken to be a linear combination of \refeq{KdVBA} evaluated on the upper and lower sheets of the Riemann surface.
\begin{equation}
	\psi_0(x,t,b,\lambda) = \psi(x,t,P^+) + b \; \psi(x,t,P^-), \quad P^\pm = (\pm \mu, \lambda).
\end{equation}

\p One may extend a little the derivation given in \cite{AGANIE} for the finite-gap field $u$ \refeq{AGKdVSol} to show that the corresponding potential $p$ which satisfies \refeq{potentialKdV} is
\begin{equation} \label{KdVFGp}
	p = -2\partial_x \log\left[\theta(Ux-4Wt+D, B)\right] - 2c_1x + 8c_3t.
\end{equation}
Then from \refeq{pq} the corresponding $q$ which solves the B{\"a}cklund equations with parameter $\sigma$ for all $x,t$ is found to be
\begin{multline} \label{KdVFGq}
	q = -2\partial_{x}\log\left[
	\theta\left(z+\int_\infty^{\sigma^+}\omega\right) + K E_\sigma \theta\left(z-\int_\infty^{\sigma^+}\omega\right)
	\right] - 2(c_1x+\Omega_1(\sigma^+)) + 8c_3t,
\end{multline}
\begin{equation}
	z = Ux-4Wt+D, \quad
	K = b_1 \frac{\theta\left(D+\int_\infty^{\sigma^+}\omega\right)}{\theta\left(D-\int_\infty^{\sigma^+}\omega\right)}, \quad
	E_\sigma = e^{-2\Omega_1(\sigma^+) x +8\Omega_3(\sigma^+) t}
\end{equation}
where the fact that for any branch point $\lambda_{BP}$
\begin{equation}
	\int_{\lambda_{BP}}^{P^-} d\Omega_{1,3} = -\int_{\lambda_{BP}}^{P^+} d\Omega_{1,3}, \qquad
	\int_{\lambda_{BP}}^{P^-} \omega = -\int_{\lambda_{BP}}^{P^+} \omega,
\end{equation}
has been used.
A solution to the defect \refeq{KdVdefectEq} placed at $x=x_D$ therefore consists of $p$ given by \refeq{KdVFGp} for $x<x_D$ and $q$ given by \refeq{KdVFGq} for $x>x_D$.

\subsection{Reality and smoothness conditions}
\label{sec:KdVReality}

For $v=q_x$ to be real it is sufficient to require that $c_1$ and the argument of the logarithm in \refeq{KdVFGq} are real.
The chosen basis of cycles is the same as for sine-Gordon but, since all branch points $\lambda_i$ are real, the relations \refeq{SGConjugate} become
\begin{equation}
	\overline{C_{ij}} = -C_{ij}, \quad
	\overline{\theta(r,B)} = \theta(\bar r,B).
\end{equation}
Therefore, $U$ and $W$ are real due to \refeq{UFormula} and \refeq{WFormula}.
The normalised holomorphic differential $\omega$ and the Abelian differentials of the second kind $d\Omega_1$, $d\Omega_3$ will all be imaginary when evaluated on a cut $[\lambda_{2n-1}, \lambda_{2n}],\ n=1\dots g$ or $[\lambda_0,\infty]$  and real otherwise maintaining $\lambda\in\real$.
This implies that the normalisation constants for $d\Omega_1$, $\alpha_i$ are real, and so, from \refeq{c1} $c_1$, will be real.

\p As was the case for sine-Gordon, in order for the theta functions in \refeq{KdVFGq} to be real each element of $\int_\infty^\sigma \omega_i$ must be real or have imaginary part $i\pi$, which requires that either $\sigma < \lambda_1$ or $\sigma$ lies in one of the intervals between cuts $[\lambda_{2k},\lambda_{2k+1}],\ k=1\dots g-1$ or $[\lambda_{2g},\lambda_{0}]$.
However, unlike sine-Gordon, $v$ will be singular at the zeros of
\begin{equation} \label{KdVSingularCondition}
	\theta\left(z+\int_\infty^{\sigma^+}\omega\right) + K E_\sigma \theta\left(z-\int_\infty^{\sigma^+}\omega\right).
\end{equation}
To ensure that \refeq{KdVSingularCondition} has no zeros for $x,t\in\real$ it is first required that $\sigma < \lambda_1$ so that $\int_{\infty}^\sigma \omega$ is real, and, since $z$ and $B$ are real, the theta functions $\theta(z\pm\int_\infty^{\sigma^+}\omega)$, as a sum of positive exponentials with real exponents, will be strictly positive.
Additionally requiring $K \geq 0$ makes \refeq{KdVSingularCondition} strictly positive.

\p If instead $\sigma$ were allowed to lie in $[\lambda_{2k},\lambda_{2k+1}],\ k=1\dots g-1$ or $[\lambda_{2g},\lambda_0]$ then one or more elements of $\int_\infty^\sigma \omega_i$ would have imaginary part $i\pi$ and the theta functions $\theta\left(z\pm\int_\infty^{\sigma^+}\omega\right)$ would have an infinite number of zeros, leading to an infinite number of singularities for $v$.  This is straightforward to see in the genus 1 case where ${\theta(i\pi + B/2 + Bm,B)=0}$ for $m\in\integers$.
If $\sigma < \lambda_1$ and $K < 0$ then $v$ would have a single singularity on an otherwise smooth background.

\p The a-periods of $d\Omega_1$ and $d\Omega_3$ are normalised to zero so by \refeq{PeriodsAsLineIntegrals} $\int_{\lambda_{2n-1}}^{\lambda_{2n}} d\Omega_1 = \int_{\lambda_{2n-1}}^{\lambda_{2n}} d\Omega_3 = 0$.
Therefore with $\sigma < \lambda_1$ the only contribution to the integrals $\Omega_1(\sigma)$ and $\Omega_3(\sigma)$ comes from the intervals $[\sigma,\lambda_1]$ and all $[\lambda_{2k},\lambda_{2k+1}],\ k=1\dots g-1$ and $[\lambda_{2g},\lambda_0]$.
But in these intervals $d\Omega_1$, $d\Omega_3$ are real so $\Omega_1(\sigma)$, $\Omega_3(\sigma)$ are real and consequently $\overline{E_\sigma} = E_\sigma$.
In conclusion, with $\sigma < \lambda_1$ and $K \geq 0$ the field to the right of the defect $v=q_x$ is real and finite.

\subsection{Soliton limit}
 It is shown in \ref{sec:partialSolitonLimit} that \refeq{KdVFGq} consists of a soliton on a finite-gap background and therefore the pair \refeq{KdVFGp} and \refeq{KdVFGq} is an algebro-geometric analogue of the solutions to the defect equations involving soliton creation on a constant background discussed in \refsec{sec:KdVsolitoncreation}.

\p To make contact with the known soliton solutions for KdV in the presence of a type I defect it is necessary to examine the soliton limit of the genus 1 finite-gap solution in which $\lambda_2 \rightarrow \lambda_1$.
Using the results of \ref{sec:FullSolitonLimit} the genus 1 finite-gap potential to the left of the defect \refeq{KdVFGp} becomes
\begin{equation}
\begin{aligned}
	p
	&= -2 \partial_{x} \log\left(1+E_1\right) + \lambda_0 x + 3\lambda_0^2 t \\
	&= - \frac{2U E_1}{1+E_1} + \lambda_0 x + 3\lambda_0^2 t.
\end{aligned}
\end{equation}
where
\begin{equation}
	E_1 = \exp\left[U(x - (U^2 - 6\lambda_0)t - x_1)\right], \quad
	U = -2i\sqrt{\lambda_1 - \lambda_0},
\end{equation}
and, before taking the soliton limit, $D$ has been defined to be $D = - x_1 U - B/2$ for some choice of $x_1 \in \real$.

\p To obtain the field to the right of the defect \refeq{KdVFGq}, let $\sigma = -\chi^2/4 + \lambda_0$ for $\chi\in\real$ and use the expressions \refeq{KdVSolitonSecondKindDiffs} for $d\Omega_1$ and $d\Omega_3$ in the soliton limit to find
\begin{align}
	\Omega_1(\sigma^+)
	&\rightarrow i\sqrt{\sigma - \lambda_0}
	= -\frac{\abs{\chi}}{2} \\
	\Omega_3(\sigma^+)
	&\rightarrow -\frac{i}{2}\sqrt{\sigma - \lambda_0}(2\sigma+\lambda_0)
	= \frac{\abs{\chi}}{4}\left(-\frac{\chi^2}{2} + 3\lambda_0\right).
\end{align}
Directly integrating \refeq{holomorphicSoliton} for the case $g=1$ gives
\begin{equation}
	\exp\left[ \int_\infty^{\sigma^+}\omega \right]
	= -\exp\left[-2i\arctan\left(\sqrt{\frac{\sigma-\lambda_0}{\lambda_0-\lambda_1}}\right) \right]
	= \frac{\abs{\chi} + U}{\abs{\chi} - U}
	= \Delta.
\end{equation}
A convenient parameterisation for $K$ will be
\begin{equation}
	K = \Delta \exp\left[-\abs{\chi} x_2\right],
	\quad x_2\in\real.
\end{equation}
This is consistent with the smoothness condition $K \geq 0$ since $\sigma < \lambda_1 < \lambda_0$ implies that $\chi=2\sqrt{\lambda_0-\sigma} > 2\sqrt{\lambda_0-\lambda_1} = U$.
Then, in the soliton limit,
\begin{equation}
	K E_\sigma \rightarrow \Delta E_2 =  \Delta \exp\left[\abs{\chi}\left(x - (\chi^2-6\lambda_0) t - x_2\right)\right].
\end{equation}
Finally, using the expressions for $c_1$ and $c_3$ in the soliton limit \refeq{KdVcSoliton}, the field to the right of the defect becomes
\begin{equation}
	q = -2\partial_{x}\log\left[
	1 + \Delta E_1 + \Delta E_2\left(1+\Delta^{-1} E_1 \right)
	\right] + \lambda_0 x + 3\lambda_0^2 t + \abs{\chi}.
\end{equation}
The expressions for $p$ and $q$ derived here as the soliton limit of the genus 1 finite-gap solutions are the same as the one-to-two soliton solution \refeq{1to2Defect} with $p_0=0$, $a\equiv U$, $c_0\equiv \lambda_0$ and $\chi>0$.

\section{Conclusion}

The principal purpose of this paper has been to establish a method to calculate the effects on quasi-periodic solutions of placing an integrable type I defect at a point $x=x_D$ on the spatial axis. Only in especially simple cases (for example, multi-soliton or genus 1 finite-gap solutions) is it possible to obtain explicit expressions for the solutions by direct substitution; in other cases, an alternative is required. This is provided by making use of B\"acklund and Darboux transformations to generate the field on the right of the defect from  the field on the left of it. By so doing, further light has been shed on the already known scattering behaviour of solitons with type I defects in the sine-Gordon and KdV systems.

\p For  finite-gap solutions, for both sine-Gordon and KdV, it was noted that, for a given choice of distinct real branch points, there is a range of values of the defect parameter for which the phase shift becomes generically complex and the reality conditions for the field to the right of the defect cannot be satisfied.
This appears to be a unique additional feature of the finite-gap solutions that has no analogue for soliton-defect scattering in either sine-Gordon or KdV.

\p Here,  only  solutions for which the defect acts as a transition between a field to the left of a defect and its B{\"a}cklund transformed partner field to the right have been considered.
This fact was critical to being able to find solutions to the defect sewing equations at a point by using solutions that would in fact satisfy the B{\"a}cklund transformation at every point.
Therefore, the fact that these solutions also happen to solve the defect sewing equations might be viewed as incidental.
However, it is not necessarily true that all solutions to the defect equations must have this property.
For example, the bound state solution for the nonlinear Schr{\"o}dinger equation in the presence of a type I defect \cite[\S 6]{Corrigan2006} only solves the defect equations for $x<x_0$ (assuming $\Omega>0$) where $x_0$ is a parameter of the solution.
It might be interesting to see how solutions that \textit{only} solve the defect equations at a point or in a region could be generated given the field on one side of the defect. It is expected that similar considerations will apply if there are several defect points though that is not explored further here.

\p It should also be noted that there exist type II integrable defects that have an auxiliary time-dependent quantity defined on the defect \cite{Corrigan2009, Robertson2014, Bowcock2016}.  Some of these, for example those permitted in sine-Gordon, may be regarded as  \lq fusions' of two type I defects and one would therefore suppose that given a field to the left of the defect the field to the right could be generated by performing two Darboux transformations and taking an appropriate limit.
However, some other equations such as the Tzitz\'{e}ica equation (also known as Bullough-Dodd-Zhiber-Shabat or $a_2^{(2)}$ affine Toda) only have type II integrable defects for which the extra degree of freedom is intrinsically necessary. This could mean that solutions might not be obtainable in the manner described above though a detailed examination of this is beyond the scope of this paper.
 On the other hand, it is worth pointing out that a defect matrix corresponding to the Tzitz\'{e}ica type II defect is given in \cite{Aguirre2011}. So, given the known finite-gap solution \cite{Cherdantsev1990} on one side of the defect, it should be possible to construct the corresponding field on the other side of the defect using a similar approach to the one taken here.
Indeed, using the method of Darboux transformations multisoliton solutions on a finite-gap background were considered in \cite{Cherdantzev1990} while a more explicit description of a soliton on an elliptic background was provided in \cite{Brezhnev1996}.

\vspace{-20pt}
 \section{Acknowledgements}

 We dedicate this article to the memory of Petr Kulish who made many important contributions to classical and quantum integrable systems. One of us (EC) is particularly indebted to him for many interesting discussions at various conferences and other meetings over a period of nearly thirty years.
RP wishes to thank the United Kingdom Engineering and Physical Sciences Research Council for financial support via a PhD studentship and EC wishes to thank the Isaac Newton Institute, Cambridge, UK and the International Institute for Theoretical Physics, Federal University of Rio Grande do Norte, Natal, Brazil, for their hospitality in January and September 2016, respectively.

 \vfill\newpage

\appendix
\section{Full soliton limit}
\label{sec:FullSolitonLimit}

It is well known that the multisoliton solutions for sine-Gordon and KdV can be recovered from the finite-gap solutions by taking pairs of branch points to coalesce \cite{Mumford, AGANIE}.
In order to facilitate checking that results obtained in the finite-gap case agree with the known results for solitons scattering with a defect it is worth repeating the argument for obtaining multisoliton solutions from finite-gap solutions.

\p For either KdV or sine-Gordon the algebraic curve for the now degenerate Riemann surface has the form
\begin{equation} \label{muSolitonLimit}
	\mu^2 = (\lambda-\lambda_0) \prod_{k=1}^{g}(\lambda-\lambda_k)^2.
\end{equation}
For KdV, $\lambda_k \in \real$, as was the case for the branch points on the original Riemann surface.
For sine-Gordon, $\lambda_0=0$ and it is also assumed that $\lambda_k \in \real$, which corresponds to all solitons being kinks or antikinks.  It is possible to obtain the breather solutions for sine-Gordon if some of the $\lambda_k$ are conjugate pairs but that aspect will not be addressed here.

\p The normalised homomorphic differentials for the degenerate Riemann surface described by \refeq{muSolitonLimit} are \cite{AGANIE}
\begin{equation} \label{holomorphicSoliton}
	\omega_k = - \frac{\sqrt{\lambda_k - \lambda_0} \prod_{\substack{l=1 \\ l \neq k}}(\lambda-\lambda_l)}{\sqrt{\lambda-\lambda_0} \prod_{j=1}^g (\lambda-\lambda_j)} d\lambda
	= - \frac{\sqrt{\lambda_k - \lambda_0}}{\sqrt{\lambda - \lambda_0}(\lambda-\lambda_k)} d\lambda,
\end{equation}
so that
\begin{equation}
	\oint_{a_j} \omega_i = 2\pi i \delta_{ij} \frac{\sqrt{\lambda_i - \lambda_0}}{\sqrt{\lambda_j-\lambda_0}}
	= 2 \pi i \delta_{ij}.
\end{equation}
The Riemann matrix is then
\begin{equation}
	B_{jk} = \oint_{b_j} \omega_k
	= 2\int_{\lambda_j}^{\lambda_0}\omega_k
	= -2\sqrt{\lambda_k-\lambda_0} \int_{\lambda_j}^{\lambda_0} \frac{d\lambda}{\sqrt{\lambda-\lambda_0}(\lambda-\lambda_k)},
\end{equation}
where the integration path from $\lambda_j$ to $\lambda_0$ is arranged to avoid all other branch points.
By definition $\lambda_i,\lambda_j<\lambda_0$ and with the assumption $\lambda_j > \lambda_k$ and making the substitution $\sqrt{\lambda-\lambda_0} = \sqrt{\lambda_k-\lambda_0} \tanh(s)$ leads to \cite{AGANIE}
\begin{align}
	B_{jk} &= -4 \,\text{arctanh} \left(\frac{\sqrt{\lambda_j-\lambda_0}}{\sqrt{\lambda_k-\lambda_0}}\right), \quad j>k \\
	B_{jk} &= B_{kj} = -4 \,\text{arctanh} \left(\frac{\sqrt{\lambda_k-\lambda_0}}{\sqrt{\lambda_j-\lambda_0}}\right), \quad j<k,
\end{align}
while for the diagonal elements $B_{ii}\rightarrow -\infty$.

\p In the limit $B_{ii}\rightarrow -\infty$ the Riemann theta function decomposes as follows \cite{Mumford}
\begin{equation} \label{thetaLimit}
\begin{aligned}
	\theta(z-\frac{B_{ii}}{2},B) &= \sum_{n\in\integers^g} e^{\frac{1}{2}nBn + \sum_i^g n_i(z_i-B_{ii}/2)} \\
	&= \sum_{n\in\integers^g} \left[ \prod_{i=1}^g e^{\frac{1}{2} n_i(n_i-1)B_{ii} + n_i z_i} \prod_{\substack{j=1 \\ j>i}}^g e^{n_i n_j B_{ij}} \right] \\
	&\xrightarrow[B_{ii} \to -\infty]{} \sum_{\substack{n=(n_1,\dots,n_g) \\ n_i = 0 \,\text{or}\, 1}} \prod_{i=1}^g e^{n_i z_i} \prod_{\substack{j=1 \\ j>i}}^g e^{n_i n_j B_{ij}} \\
	&= 1 + \sum_{i=1}^g e^{z_i} + \sum_{\substack{j=1 \\ j>i}}^g e^{z_i + z_j} e^{B_{ij}} + \sum_{\substack{k=1 \\ k>j>i}}^g e^{z_i + z_j + z_k} e^{B_{ij} + B_{ik} + B_{jk}} + \cdots.
\end{aligned}
\end{equation}

\subsection{KdV}

This subsection is devoted to the soliton limit specifically for KdV.
Comparing \refeq{holomorphicSoliton} with \refeq{holomorphicDiff} gives
\begin{equation}
	C_{ig} = -\sqrt{\lambda_i - \lambda_0}, \qquad
	C_{i(g-1)} = \sqrt{\lambda_i - \lambda_0} \sum_{\substack{l=1 \\ l \neq i}}^g \lambda_l,
\end{equation}
so that, on using \refeq{UFormula} and \refeq{WFormula},
\begin{align}
	U_i &= -2i\sqrt{\lambda_i - \lambda_0} \label{USoliton} \\
	W_i &= 2i\sqrt{\lambda_i - \lambda_0} \left(\lambda_i + \frac{\lambda_0}{2}\right).
\end{align}
The differentials of the second kind become
\begin{align}
	d\Omega_1 &= \frac{i\lambda^g d\lambda}{2\sqrt{\lambda-\lambda_0} \prod_{j=1}^g (\lambda-\lambda_j)}
	+ \sum_{i=1}^g \alpha_i \omega_i \\
	d\Omega_3 &= -\frac{3i}{4} \left( \frac{2\lambda^{g+1} - \widehat{E} \lambda^g}{\sqrt{\lambda-\lambda_0} \prod_{j=1}^g (\lambda-\lambda_j)} \right) d\lambda + \sum_{i=1}^g \beta_i \omega_i,
	\quad \widehat{E} = \lambda_0 + 2\sum_{i=1}^{g} \lambda_i,
\end{align}
so that the normalisation constants defined by \refeq{KdVsecondNormalization} are now
\begin{align}
	\alpha_i &= \frac{i\lambda_i^g}{2\sqrt{\lambda_i-\lambda_0} \prod_{\substack{k=1 \\ k \neq i}}^g (\lambda_i - \lambda_k)} \\
	\beta_i &= -\frac{3i}{4} \left(\frac{2\lambda_i^{g+1} - \widehat{E} \lambda_i^g}{\sqrt{\lambda_i-\lambda_0} \prod_{\substack{k=1 \\ k \neq i}}^g (\lambda_i - \lambda_k)}\right).
\end{align}
This simplifies $d\Omega_1$ and $d\Omega_3$ to
\begin{equation} \label{KdVSolitonSecondKindDiffs}
	d\Omega_1 = \frac{i}{2\sqrt{\lambda-\lambda_0}}, \qquad
	d\Omega_3 = -\frac{3i(2\lambda - \lambda_0)}{4\sqrt{\lambda-\lambda_0}}.
\end{equation}

\p The coefficients $c_1$ and $c_3$ in the asymptotic expansions \refeq{KdVSecondKindExpansion} can then be found, either by directly expanding the integrals $\Omega_1$ and $\Omega_3$ or by using the formulas \refeq{c1} and \refeq{c3}, to be
\begin{equation} \label{KdVcSoliton}
	c_1 = -\frac{\lambda_0}{2}, \qquad  c_3 = \frac{3\lambda_0^2}{8}.
\end{equation}

\p In the soliton limit the potential $p$ \refeq{KdVFGp}, corresponding to the field $u=p_x$, which solves KdV becomes,
\begin{gather}
	p(x,t) = -2 \partial_{x} \log\left[\sum_{\substack{n=(n_1,\dots,n_g) \\ n_i = 0 \,\text{or}\, 1}} \prod_{i=1}^g e^{n_i z_i(x,t)} \prod_{\substack{j=1 \\ j>i}}^g e^{n_i n_j B_{ij}}\right] + \lambda_0 x + 3\lambda_0^2 t \label{multisolitonKdV} \\
	z_i(x,t) = U_i(x - (U_i^2 - 6\lambda_0)t -(x_0)_i), \label{multisoliton_zi}
\end{gather}
where $D_i = - (x_0)_i U_i - B_{ii}/2$ for some choice of $x_0 \in \real^g$ and the limit of the theta function \refeq{thetaLimit} has also been used.
This is a $g$-soliton solution with each $B_{ij}$ corresponding to the interaction of the $i$th and $j$th solitons.

\subsection{sine-Gordon}
\label{SGSolitonLimit}

In the sine-Gordon case $\lambda_0=0$ and therefore $\lambda_k<0$ for $k=1,\dots,g$.  The normalisation constants are then
\begin{equation}
	C_{ig} = -\sqrt{\lambda_i}, \qquad
	C_{i1} = -\sqrt{\lambda_i} \prod_{\substack{k=1 \\ k \neq i}}^g (-\lambda_k),
\end{equation}
and so from \refeq{SGSecondB}
\begin{equation}
	(B_1)_i = -2\sqrt{\lambda_i}, \qquad
	(B_2)_i = -\frac{2}{\sqrt{\lambda_i}}.
\end{equation}

\p The Abelian differentials of the second kind $d\Omega_1$, $d\Omega_2$ in the soliton limit become
\begin{align}
	d\Omega_1 &= \frac{\lambda^g d\lambda}{2\sqrt{\lambda} \prod_{j=1}^g (\lambda-\lambda_j)} + \sum_{i=1}^g \alpha_i \omega_i ,\\
	d\Omega_2 &= (-1)^{g+1} \frac{\prod_{k=1}^g \lambda_k}{2\lambda\sqrt{\lambda} \prod_{j=1}^g (\lambda-\lambda_j)} d\lambda + \sum_{i=1}^g \beta_i \omega_i,
\end{align}
so their respective normalisation constants are
\begin{align}
	\alpha_i &= \frac{\lambda_i^g}{2\sqrt{\lambda_i} \prod_{\substack{j=1 \\ j \neq i}}^g (\lambda_i-\lambda_j)}, \\
	\beta_i &= (-1)^{g+1} \frac{\prod_{k=1}^g \lambda_k}{2\lambda_i \sqrt{\lambda_i} \prod_{\substack{j=1 \\ j \neq i}}^g (\lambda_i-\lambda_j)}.
\end{align}
This simplifies the differentials to
\begin{equation} \label{SGSecondKindDiff_SolitonLimit}
	d\Omega_1 = \frac{1}{2\sqrt{\lambda}}, \qquad
	d\Omega_2 = -\frac{1}{2\lambda\sqrt{\lambda}}.
\end{equation}

\p Turning now to the limit of the  genus $g$ solution to the sine-Gordon equation \refeq{SGfinitegap}, define the rapidity,
\begin{equation}
	\theta_i = -\frac{1}{2} \log(-\lambda_i)
\end{equation}
and let
\begin{equation}
	D_i = (x_0)_i + \frac{i\pi}{2} + i\pi\varepsilon_i - \frac{B_{ii}}{2},
\end{equation}
for a choice of $x_0 \in \real^g$ and each $\varepsilon_i = 0$ or $1$.
Then \refeq{SGfinitegap} becomes
\begin{gather}
	\exp\left(\frac{i u(x,t)}{2}\right) = \frac{\tau_0}{\tau_1} \\
	\tau_\alpha = \sum_{\substack{n=(n_1,\dots,n_g) \\ n_i = 0 \,\text{or}\, 1}} \prod_{i=1}^g \left[i (-1)^{\varepsilon_i+\alpha} e^{z_i(x,t)}\right]^{n_i} \prod_{\substack{j=1 \\ j>i}}^g e^{n_i n_j B_{ij}} \label{SGmultisoliton} \\
	z_i(x,t) = \cosh\theta_i\, x - \sinh\theta_i\, t + (x_0)_i, \label{SGmultisoliton_zi}
\end{gather}
 which is then recognised to be the multisoliton solution for sine-Gordon written in the Hirota form \cite{Hirota1972}.

\section{Partial soliton limit}
\label{sec:partialSolitonLimit}

\begin{figure}
	\center
	\includegraphics[width=15cm]{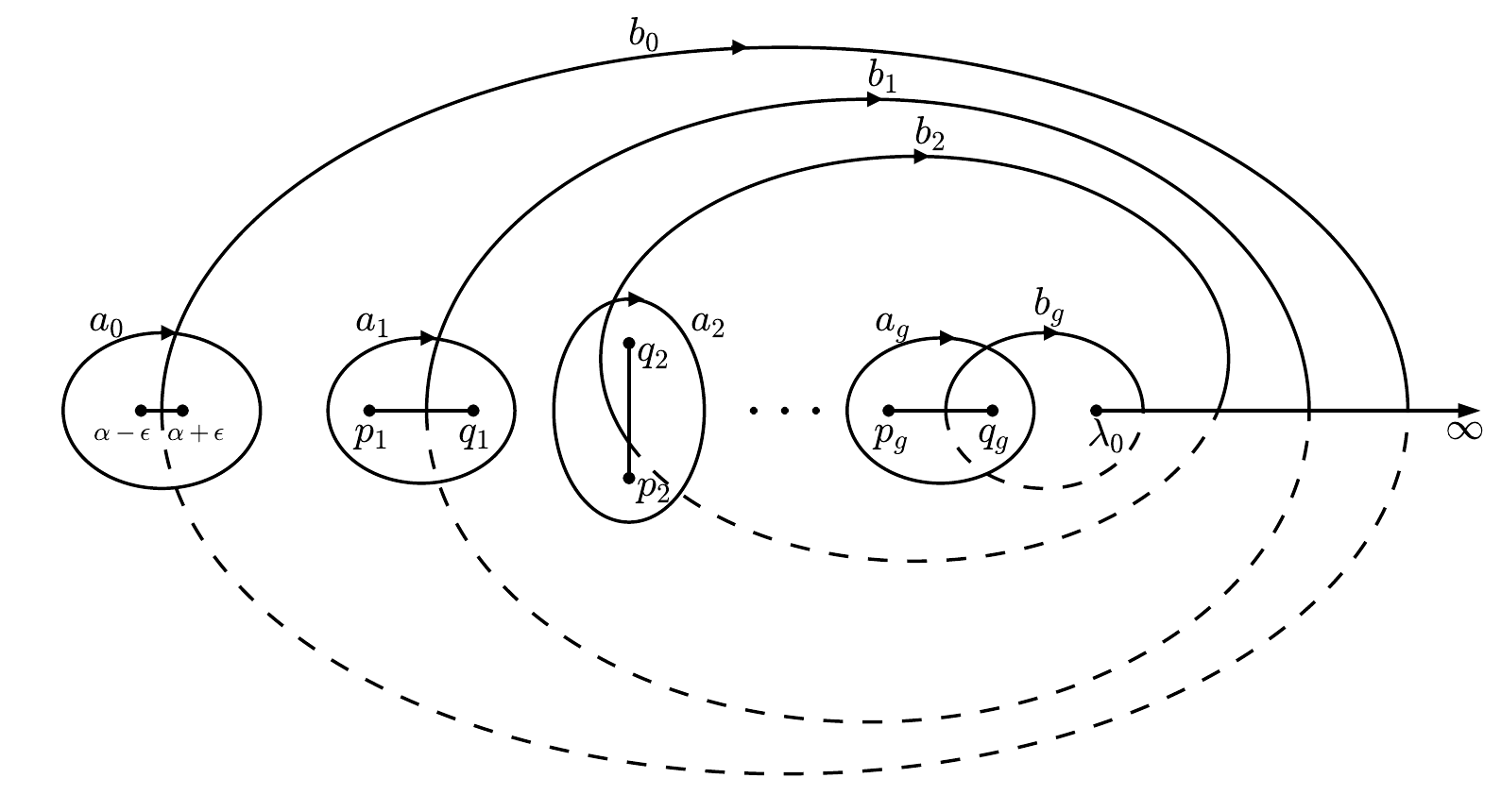}
	\caption{
	The chosen basis of cycles for a  genus $g+1$ Riemann surface which in the limit $\epsilon\rightarrow 0$ becomes the partially degenerate surface $\tilde X$.
	}
	\label{PartialHomology}
\end{figure}

\p Certain integrals on the Riemann surface $X$ of genus $g$ defined by the algebraic curve
\begin{equation}
	\mu^2(\lambda) = (\lambda-\lambda_0) \prod\limits_{i=1}^{g}(\lambda-p_i)(\lambda-q_i)
\end{equation}
need to be related to constants associated with the partially degenerate Riemann surface $\tilde X$, which is the same as $X$ except for a singularity at $\alpha$.
In other words, think of $\tilde X$ as being the limit of a genus $g+1$ surface in which one pair of branch points coalesces to $\alpha$ and the remaining branch points are the same as for $X$, as shown in \reffig{PartialHomology}.
Associated with this coalescing pair of branch points the cycles $a_0$ and $b_0$ are defined analogously to the other cycles, as shown in \reffig{PartialHomology}.
As the pair of branch points coalesce, $\tilde B_{00} \rightarrow -\infty$ and the theta function associated with $\tilde X$ becomes
\begin{equation}
	\theta\left(z_\mu - \delta_{\mu0} \frac{\tilde B_{00}}{2}
,\tilde B_{\mu \nu}\right) \rightarrow \theta(z_i,\tilde B_{ij}) + \theta(z_{i} + \tilde B_{0i},\tilde B_{ij}) e^{z_0},
\end{equation}
where here and throughout this section the indices $i,j$ run from $1$ to $g$ and $\mu,\nu$ from $0$ to $g$ unless otherwise stated.

\p The finite-gap solution to sine-Gordon corresponding to the surface $\tilde X$ would then be, from \refeq{SGfinitegap},
\begin{equation} \label{SGPartiallyDegenerateSol}
	e^{iv/2} = \frac{\theta(z_i,\tilde B_{ij}) + \theta(z_{i} + \tilde B_{0i},\tilde B_{ij}) e^{z_0}}{\theta(z_i + i\pi,\tilde B_{ij}) - \theta(z_{i} + i\pi + \tilde B_{0i},\tilde B_{ij}) e^{z_0}},
\end{equation}
where $z_\mu = i\rho (\tilde B_1)_\mu + i\xi (\tilde B_2)_\mu + D_\mu$ and the $\tilde B_{00}/2$ has been absorbed into the constant $D_0$.
The corresponding KdV solution from \refeq{AGKdVSol} would be
\begin{equation} \label{KdVPartiallyDegenerateSol}
	v = -2\partial_{xx}\log\left[\theta(z_i,\tilde B_{ij}) + \theta(z_{i} + \tilde B_{0i},\tilde B_{ij}) e^{z_0}\right] -2\tilde c_1,
\end{equation}
where $z_\mu = \tilde U_\mu x - 4\tilde W_\mu t + D_\mu$ and again $\tilde B_{00}/2$ has been absorbed into $D_0$.

\p The limit in which one or more branch points coalesce is known to lead to soliton solutions \cite{Mumford, AGANIE} (see also \ref{sec:FullSolitonLimit}) so the solutions \refeq{SGPartiallyDegenerateSol} and \refeq{KdVPartiallyDegenerateSol} are one soliton solutions on a genus $g$ finite-gap background.
 It is necessary to show that the field to the right of the defect for sine-Gordon \refeq{SGDarbouxTransformedFiniteGap} and KdV \refeq{KdVFGq} that was found through the Darboux transformation of the given finite-gap field to the left has the same form as \refeq{SGPartiallyDegenerateSol} and \refeq{KdVPartiallyDegenerateSol}, respectively.

\p The holomorphic differentials $\omega$, $\tilde\omega$ associated with the surfaces $X$, $\tilde X$ have the form
\begin{align}
	\omega_i &= \frac{\varphi_i(\lambda)}{\mu(\lambda)} d\lambda, \qquad
	&& \varphi_i(\lambda) = \sum_{j=1}^g C_{ij}\lambda^{j-1}
	&&& i=1,\dots,g \\
	\tilde\omega_\mu &= \frac{\tilde\varphi_\mu(\lambda)}{\mu(\lambda)(\lambda-\alpha)} d\lambda, \qquad
	&& \tilde\varphi_\mu(\lambda) = \sum_{\nu=0}^g \tilde C_{\mu\nu}\lambda^{\nu}
	&&& \mu=0,\dots,g.
\end{align}

\p The normalisation condition
\begin{equation}
	\oint_{a_0}\tilde\omega_\mu
	= -2\pi i \frac{\tilde\varphi_\mu(\alpha)}{\mu(\alpha)}
	= 2\pi i \delta_{0\mu},
\end{equation}
implies that
\begin{gather}
	\tilde\varphi_0(\lambda)
	= -\mu(\lambda) + \sum_{j=1}^{g} \tilde C_{0j} (\lambda^j - \alpha^j) \\
	\tilde\varphi_{i}(\lambda)
	= \sum_{j=1}^{g} \tilde C_{ij} (\lambda^j - \alpha^j), \quad i=1,\dots,g.
\end{gather}

\p Since
\begin{equation*}
	\sum_{j=1}^g \tilde C_{ij} (\lambda^j-\alpha^j)
	= (\lambda-\alpha) \sum_{j=1}^g \tilde C_{ij} \sum_{k=1}^j \alpha^{j-k}\lambda^{k-1}
	= (\lambda-\alpha) \sum_{k=1}^g \lambda^{k-1} \sum_{j=k}^g \tilde C_{ij} \alpha^{j-k}
\end{equation*}
the holomorphic differential for the partially degenerate surface has the form
\begin{gather}
	\tilde\omega_0 = -\frac{\mu(\alpha)}{\mu(\lambda)(\lambda-\alpha)} d\lambda
	+ \sum_{k=1}^g \sum_{j=k}^g \frac{\tilde C_{0j} \alpha^{j-k} \lambda^{k-1}}{\mu(\lambda)} d\lambda \label{tildeomega0} \\
	\tilde\omega_{i} = \sum_{k=1}^g \sum_{j=k}^g \frac{\tilde C_{ij} \alpha^{j-k} \lambda^{k-1}}{\mu(\lambda)} d\lambda, \quad i=1,\dots,g
\end{gather}
But $\tilde\omega_{i>0}$ is of the same form as $\omega_i$ and they have the same normalisation condition for the same cycles $a_{i>0}$
\begin{equation}
	\oint_{a_{i}} \tilde\omega_{j} = 2\pi i \delta_{ij}, \qquad
	\oint_{a_i} \omega_j = 2\pi i \delta_{ij}, \quad i,j=1,\dots,g
\end{equation}
and so
\begin{equation} \label{CtildeC}
	\omega_i = \tilde\omega_i , \qquad C_{ik} = \sum_{j=k}^g \tilde C_{ij} \alpha^{j-k}, \quad i,k=1,\dots,g
\end{equation}

\p It is now clear that $\tilde B_{ij} = B_{ij}$ for $i,j=1,\dots,g$ and
\begin{equation} \label{tildeB0i}
	\tilde B_{0i}
	= \oint_{b_0} \tilde\omega_i
	= 2\left[\int_{\alpha}^{p_1} + \sum_{k=1}^{g-1} \int_{q_k}^{p_{k+1}} + \int_{q_g}^{\lambda_0} \right] \tilde\omega_i
	= -2\int_\infty^{\alpha} \tilde \omega_i
	= -2\int_\infty^{\alpha^+} \omega_i
\end{equation}
for $i=1,\dots,g$ and where in the second equality  \refeq{PeriodsAsLineIntegrals} has been used and in the third equality  \refeq{aCycleSum} has been used to add
\begin{equation}
-\int_{\infty}^{\lambda_0} \tilde\omega_j + \sum_{k=1}^g \int_{p_k}^{q_k} \tilde \omega_j = 0
\qquad j=1,\dots,g.
\end{equation}

\subsection{sine-Gordon}
Turning now to the parameters specifically associated with sine-Gordon the differential of the second kind for the partially degenerate surface $\tilde X$ is
\begin{equation}
	d\tilde\Omega_1 = \frac{\lambda^{g+1}}{2\mu(\lambda)(\lambda-\alpha)} d\lambda + \sum_{\mu=0}^g \tilde\alpha_\mu \tilde\omega_\mu,
\end{equation}
with the normalisation condition
\begin{equation}
	\oint_{a_0} d\tilde\Omega_1 = 0,
\end{equation}
which implies that
\begin{equation}
	\tilde\alpha_0 = \frac{\alpha^{g+1}}{2\mu(\alpha)}.
\end{equation}
So, taking into account $\tilde\omega_{i>0} = \omega_i$ and using \refeq{tildeomega0},
\begin{align}
	d\tilde\Omega_1
	&= \frac{\lambda^{g+1}-\alpha^{g+1}}{2\mu(\lambda)(\lambda-\alpha)} d\lambda +
	\sum_{k=1}^g \left[\tilde\alpha_0 \sum_{j=k}^g \tilde C_{0j} \alpha^{j-k} + \sum_{i=1}^g \tilde\alpha_i C_{ik} \right] \frac{\lambda^{k-1}}{\mu(\lambda)} d\lambda \\
	&= \sum_{k=1}^{g+1} \frac{\alpha^{g+1-k}\lambda^{k-1}}{2\mu(\lambda)} d\lambda +
	\sum_{k=1}^g \left[\tilde\alpha_0 \sum_{j=k}^g \tilde C_{0j} \alpha^{j-k} + \sum_{i=1}^g \tilde\alpha_i C_{ik} \right] \frac{\lambda^{k-1}}{\mu(\lambda)} d\lambda \\
	&= \frac{\lambda^g}{2\mu(\lambda)} d\lambda + \sum_{k=1}^g \left[\frac{\alpha^{g+1-k}}{2} + \tilde\alpha_0 \sum_{j=k}^g \tilde C_{0j} \alpha^{j-k} + \sum_{i=1}^g \tilde\alpha_i C_{ik} \right] \frac{\lambda^{k-1}}{\mu(\lambda)} d\lambda.
\end{align}
This is of the same form as the corresponding differential for $X$,
\begin{equation}
	d\Omega_1 = \frac{\lambda^g}{2\mu(\lambda)}d\lambda + \sum_{i=1}^g \sum_{k=1}^g \alpha_i C_{ik} \lambda^{k-1},
\end{equation}
and, because the remaining normalisation conditions are the same,
\begin{equation}
	\oint_{a_i} d\tilde\Omega_1 = 0 \qquad \oint_{a_i} d\Omega_1 = 0 \qquad i=1 \dots g,
\end{equation}
it follows that
\begin{equation}
	d\tilde \Omega_1 = d\Omega_1, \qquad
	\sum_{i=1}^g \alpha_i C_{ik} = \frac{\alpha^{g+1-k}}{2} + \tilde\alpha_0 \sum_{j=k}^g \tilde C_{0j} \alpha^{j-k} + \sum_{i=1}^g \tilde\alpha_i C_{ik}.
\end{equation}
Therefore for the $b_{i>0}$ periods,
\begin{equation}
	(\tilde B_1)_i = \oint_{b_i} d\tilde\Omega_1 = \oint_{b_i} d\Omega_1 = (B_1)_i \qquad i=1 \dots g,
\end{equation}
which could also have been derived using \refeq{SGSecondB} and \refeq{CtildeC}.
Writing the $b_0$ cycle as the sum of line integrals as in \refeq{tildeB0i}
\begin{equation}
	(\tilde B_1)_0
	= \oint_{b_0} d\tilde\Omega_1
	= 2\left[\int_{\alpha}^{p_1} + \sum_{k=1}^{g-1} \int_{q_k}^{p_{k+1}} + \int_{q_g}^{0} \right] d\tilde\Omega_1
	= -2\int_0^{\alpha} d\tilde\Omega_1
	= -2\int_0^{\alpha^+} d\Omega_1,
\end{equation}
where in the third equality the normalisation conditions have been used to add
\begin{equation}
	0 = \oint_{a_i} d\tilde\Omega_1 = \int_{p_i}^{q_i} d\tilde\Omega_1.
\end{equation}

\p Treating the other differential of the second kind in the same manner
\begin{equation}
	d\tilde\Omega_2 = -\frac{\sqrt{\Lambda \alpha^2}}{2\lambda\mu(\lambda)(\lambda-\alpha)} d\lambda + \sum_{\mu=0}^g \tilde\beta_\mu \tilde\omega_\mu, \qquad \Lambda = \prod_{i=1}^g p_i q_i,
\end{equation}
the normalisation condition for the $a_0$ cycles gives
\begin{equation}
	\oint_{a_0} d\tilde\Omega_2 = 0 \quad \implies \quad \tilde\beta_0 = -\frac{\sqrt{\Lambda \alpha^2}}{2\alpha\mu(\alpha)}.
\end{equation}
Assuming $\alpha < 0$, since in this application $\alpha = -\sigma^2$ and $\sigma>0$,
\begin{equation}
	d\tilde\Omega_2 = -\frac{\sqrt{\Lambda}}{2\lambda\mu(\lambda)} d\lambda + \sum_{k=1}^g \left[ \tilde\beta_0 \sum_{j=k}^g \tilde C_{0j} \alpha^{j-k} + \sum_{i=1}^g \tilde \beta_i C_{ik} \right] \frac{\lambda^{k-1}}{\mu(\lambda)},
\end{equation}
which is of the same form as \refeq{SGsecondKindDiff2} and therefore
\begin{equation}
	d\tilde \Omega_2 = d\Omega_2, \qquad
	\sum_{i=1}^g \beta_i C_{ik} = \tilde\beta_0 \sum_{j=k}^g \tilde C_{0j} \alpha^{j-k} + \sum_{i=1}^g \tilde \beta_i C_{ik}.
\end{equation}
Decomposing the $b$ periods and taking into account the same normalisation condition gives
\begin{equation}
	(\tilde B_2)_0
	= \oint_{b_0} d\tilde\Omega_2
	= 2\left[\int_{\alpha}^{p_1} + \sum_{k=1}^{g-1} \int_{q_k}^{p_{k+1}} + \int_{q_g}^{\infty} \right] d\tilde\Omega_2
	= -2\int_\infty^{\alpha} d\tilde\Omega_2
	= -2\int_\infty^{\alpha^+} d\Omega_2.
\end{equation}

\p Returning to \refeq{SGPartiallyDegenerateSol}  the solution to the sine-Gordon equation corresponding to the partially degenerate surface $\tilde X$ is now
\begin{equation}
\begin{gathered} \label{SGPartiallyDegenerateSolFinal}
	e^{iv/2} = \frac{\theta(z,B) + \theta\left(z - 2\int_\inf^{\alpha^+} \omega ,B\right) e^{z_0}}{\theta(z + i\pi,B) - \theta\left(z + i\pi - 2\int_\inf^{\alpha^+} \omega, B\right) e^{z_0}} \\
	z = iB_1 \rho + iB_2 \xi + D, \qquad z_0 = -2i\left(\rho \int_0^{\alpha^+} d\Omega_1 + \xi \int_\inf^{\alpha^+} d\Omega_2 \right) + D_0.
\end{gathered}
\end{equation}

\p In the above expression $\int_\inf^{\alpha^+} \omega$ can be replaced with $\int_0^{\alpha^+} \omega$ since
\begin{equation*}
	\int_\infty^{\alpha^+} \omega = \int_\infty^0 \omega + \int_0^{\alpha^+} \omega = i\pi + \int_0^{\alpha^+} \omega,
\end{equation*}
and the Riemann theta function is periodic in $2\pi i$.
Then after letting $\alpha = -\sigma^2$ and shifting $D \rightarrow D + \int_0^{-\sigma^2} \omega$ it can be seen that the one soliton solution on a  genus $g$ background \refeq{SGPartiallyDegenerateSolFinal} is of the same form as the field to the right of the defect \refeq{SGDarbouxTransformedFiniteGap} obtained via a Darboux transformation of the original  genus $g$ finite-gap field to the left of the defect, where
\begin{equation*}
	e^{D_0} = b \frac{\theta\left(D + \int_0^{-\sigma^2} \omega, B\right)}{\theta\left(D - \int_0^{-\sigma^2} \omega, B\right)}.
\end{equation*}

\subsection{KdV}

For KdV the procedure is similar.  The differentials of the second kind for the partially degenerate surface $\tilde X$ are
\begin{align}
	d\tilde\Omega_1 &= \frac{i\lambda^{g+1}}{2\mu(\lambda)(\lambda-\alpha)} d\lambda + \sum_{\mu=0}^g \tilde\alpha_\mu \tilde\omega_\mu, \\
	d\tilde\Omega_3 &= -\frac{3i}{4} \left( \frac{2\lambda^{g+2} - (2\alpha + \widehat{E}) \lambda^{g+1}}{\mu(\lambda)(\lambda-\alpha)} \right) d\lambda + \sum_{\mu=0}^g \tilde\beta_i \tilde\omega_i, \quad
	\widehat{E} = \sum_{i=0}^{2g} \lambda_i.
\end{align}

\p The normalisation conditions
\begin{equation}
	\oint_{a_0} d\tilde\Omega_1 = 0, \qquad
	\oint_{a_0} d\tilde\Omega_3 = 0,
\end{equation}
imply that
\begin{equation}
	\tilde\alpha_0 = \frac{i\alpha^{g+1}}{2\mu(\alpha)}, \qquad
	\tilde\beta_0 = \frac{3i\widehat{E}\alpha^{g+1}}{4\mu(\alpha)}.
\end{equation}
Using the explicit form of $\tilde\omega_0$, \refeq{tildeomega0}, and the fact that $\tilde\omega_{i>0} = \omega_i$, the differentials for $\tilde X$ and their normalisation constants can be related to their counterparts for $X$:
\begin{align}
	d\tilde\Omega_1
	&= d\Omega_1, \quad
	&& \sum_{i=1}^g \alpha_i C_{ik} = \frac{i}{2}\alpha^{g+1-k} + \tilde\alpha_0 \sum_{j=k}^g \tilde C_{0j} \alpha^{j-k} + \sum_{i=1}^g \tilde\alpha_i C_{ik}
	\\
	d\tilde\Omega_3
	&= d\Omega_3, \quad
	&& \sum_{i=1}^g \beta_i C_{ik} = \frac{3i\widehat E}{4}\alpha^{g+1-k} + \tilde\beta_0 \sum_{j=k}^g \tilde C_{0j} \alpha^{j-k} + \sum_{i=1}^g \tilde\beta_i C_{ik}.
\end{align}
It  follows immediately that the periods around $b_{i>0}$ for the differentials of the second kind are therefore the same for both surfaces
\begin{align}
	\tilde U_i &= \oint_{b_i} d\tilde\Omega_1 = \oint_{b_i} d\Omega_1 = U_i \qquad i=1 \dots g \\
	\tilde W_i &= \oint_{b_i} d\tilde\Omega_3 = \oint_{b_i} d\Omega_3 = W_i,
\end{align}
and that the coefficients of $k^{-1}$ at $\infty$ for the two differentials are the same on both surfaces,
\begin{equation}
	\tilde c_1 = c_1, \qquad \tilde c_3 = c_3.
\end{equation}

\p The period around $b_0$ for the differentials on $\tilde X$ can be written as in integral over $X$ by using \refeq{PeriodsAsLineIntegrals} to decompose the cycle into line integrals
\begin{align*}
	\tilde U_0
	= \oint_{b_0} d\tilde\Omega_1
	= 2\left[\int_{\alpha}^{\lambda_1} + \sum_{k=1}^{g-1} \int_{\lambda_{2k}}^{\lambda_{2k+1}} + \int_{\lambda_{2g}}^{\lambda_0} \right] d\tilde\Omega_1
	= -2\int_{\lambda_0}^\alpha d\tilde\Omega_1
	= -2\int_{\lambda_0}^{\alpha^+} d\Omega_1
	\\
	\tilde W_0
	= \oint_{b_0} d\tilde\Omega_3
	= 2\left[\int_{\alpha}^{\lambda_1} + \sum_{k=1}^{g-1} \int_{\lambda_{2k}}^{\lambda_{2k+1}} + \int_{\lambda_{2g}}^{\lambda_0} \right] d\tilde\Omega_3
	= -2\int_{\lambda_0}^\alpha d\tilde\Omega_3
	= -2\int_{\lambda_0}^{\alpha^+} d\Omega_3,
\end{align*}
and in the third equality within each line the following have been used
\begin{equation}
	\int_{\lambda_{2i-1}}^{\lambda_{2i}} d\tilde\Omega_1 = \oint_{a_i} d\tilde\Omega_1 = 0, \qquad
	\int_{\lambda_{2i-1}}^{\lambda_{2i}} d\tilde\Omega_3 = \oint_{a_i} d\tilde\Omega_3 = 0.
\end{equation}

\p After setting $\alpha = \sigma$ the one soliton solution on a genus $g$ background \refeq{KdVPartiallyDegenerateSol} corresponding to the partially degenerate surface $\tilde X$ can now be written as
\begin{gather}
	v = -2\partial_{xx}\log\left[\theta(z_i,B_{ij}) + \theta\left(z_i - 2\int_{\inf}^{\sigma^+}\omega_i,B_{ij}\right) e^{z_0}\right] -2c_1 \\
	 z_i = U_i x - 4W_i t + D_i, \qquad
	  z_0 = -2x\int_{\lambda_0}^{\sigma^+}d\Omega_1  + 8t \int_{\lambda_0}^{\sigma^+}d\Omega_3 + D_0,
\end{gather}
which, after shifting $D_i \rightarrow D_i + \int_{\inf}^{\sigma^+}\omega_i$, (the reality conditions \refsec{sec:KdVReality} already require $\int_{\inf}^{\sigma^+}\omega_i \in\real$), has the same form as the field $v=q_x$ where $q$ \refeq{KdVFGq} was obtained by the Darboux transformation of the original genus $g$ potential $p$ \refeq{KdVFGp} corresponding to the surface $X$.

\raggedright

\end{document}